\documentclass[structabstract]{aa}
\usepackage{graphicx}                %  Include graphics
\usepackage{txfonts}                 %  Postscript Fonts

\newcommand{\kms}{km\,s$^{-1}$}
\newcommand{\fok}{\hbox{$^\circ$}}

\begin{document}

\title{Multiple and changing cycles of active stars}
\subtitle{II. Results}

\author{K. Ol\'ah\inst{1} 
\and Z. Koll\'ath\inst{1} 
\and T. Granzer\inst{2} \and  K.G. Strassmeier\inst{2} 
\and A.F. Lanza \inst{5} 
\and S. J\"arvinen\inst{2,6,7} 
\and H. Korhonen\inst{3},
\newline S.L. Baliunas\inst{4} \and W. Soon\inst{4} \and S. Messina\inst{5} \and 
G. Cutispoto\inst{5}}

\offprints{K. Ol\'ah}

\institute{Konkoly Observatory of the Hungarian Academy of Sciences, H-1525 Budapest, Hungary; \email olah@konkoly.hu 
\and Astrophysical Institute Potsdam (AIP), An der Sternwarte 16, D-14482 Potsdam, Germany 
\and ESO, Karl-Schwarzschild Strasse 2, 85748 Garching bei Munchen, Germany  
\and Harvard--Smithsonian Center for Astrophysics, Cambridge, MA 02138, USA 
\and INAF - Osservatorio Astrofisico di Catania, via S. Sofia 78, 95123 Catania, Italy
\and Tuorla Observatory, University of Turku, 21500 Piikki\"o, Finland
\and Astronomy Division, P.O. Box 3000, 90014 University of Oulu, Finland
}
\date{Received ; accepted}
\authorrunning{Ol\'ah et al.}
\titlerunning{Changing cycles -- results}

\abstract{}{We study the time variations of the cycles of 20 active stars based 
on decades-long photometric or spectroscopic observations.}{A method of 
time-frequency analysis, as discussed in a companion paper, is applied to the 
data.}{Fifteen stars definitely show multiple cycles; the records of the rest 
are too short to verify a timescale for a second cycle. The cycles typically 
show systematic changes. For three stars, we found two cycles in each of them that 
are not harmonics, and which vary in parallel, indicating that a common physical 
mechanism arising from a dynamo construct. The positive relation between the 
rotational and cycle periods is confirmed for the inhomogeneous set of active 
stars.}{Stellar activity cycles are generally multiple and variable.}


\maketitle

%%%%%%%%%%%%%%%%%%%%%%%%%%%%%%%%%
%%%%%%%%% I N T R O %%%%%%%%%%%%%
%%%%%%%%%%%%%%%%%%%%%%%%%%%%%%%%%

\section{Introduction}

In the middle of the last century it was realised that certain observed features 
of late-type stars could be explained by magnetic phenomena (Kron \cite{kron}), similar to those 
that had been detected on the Sun.
The cyclic pattern of solar activity had already been known for more than 120 
years (Schwabe \cite{schwabe}) by the time systematic research of magnetic 
activity of late-type stars commenced. In 1966 long-term monitoring of the
relative fluxes of cores of the CaII lines -- thought to indicate the
strength and coverage of surface magnetism through the enhancement
of chromospheric flux in the line cores -- of solar type stars began (cf. Wilson 
\cite{wilson_old}), to search for stellar cycles analogous to the solar case, 
and that monitoring program continued for more than three decades. Eleven years 
after systematic monitoring had begun (Wilson \cite{wilson}) published landmark
results on the first decadal survey of chromospheric variability and stellar 
magnetic cycles.

While Wilson was early in his observational work, a contemporary comment on the
possibility of the existence of stellar cycles appeared in an editorial note of 
Detre (\cite{detre}) to a paper presenting photometric observations of BY~Dra by 
P.F. Chugainov, as follows: {\it The continuous observation of Chugainov's stars 
would be extremely important because diagrams like that on the opposite page may 
reveal the existence of cycles similar to the solar cycle in these stars.}

Studying the late type eclipsing binary XY~UMa, Geyer (\cite{geyer}) suggested 
an activity cycle to explain the outside-eclipse variability of the system. 
Durney \& Stenflo (\cite{d_s}) presented a first theoretical forecast, namely,  
with increasing rotation rate, cycle periods should be decreasing. Baliunas et al.
(\cite{sallie4}) summarized knowledge on the dynamo interpretation of stellar 
activity cycles prior to 1996.

As databases relevant to decadal stellar activity continued to lengthen and 
contain more stars, study began on the variability of stellar activity cycles. 
Multiple cycles, analogous to the known solar multicyclic variability, were 
recovered by Baliunas et al. (\cite{sallie4}) from the records of the Ca II index. 
Using these Ca index data supplemented with photometric results of active dwarfs and giants Saar \& Brandenburg (\cite{saar}) studied time variability of stellar cycles and multiple cycles on the evolutionaly timescale, draw the distribution of $P_{rot}/P_{cyc}$ in the function of the Rossby number and compared the results with theoretical models.

At the beginning of the current century, several photometric records became long 
enough to study photospheric magnetic cycles on a relatively large sample of 
active stars. Using conventional methods Ol\'ah et al. (\cite{cycles1}, \cite 
{cycles2}) derived activity cycles and multicycles for most objects in the 
present investigation, and Messina \& Guinan (\cite{messina}) for six young 
solar analogues. Radick et al. (\cite{radick}) investigated the relation between 
the Ca index and contemporaneous photometric measurements on 35 Sun-like stars. 
Lockwood et al. (\cite{lockwood}) studied the magnetic cycle patterns seen in 
the photospheres and chromospheres of 32 stars primarily on or near the lower 
main-sequence. In both those last two papers it was found that on a decadal 
scale, younger stars decrease in brightness when their chromosperic activity 
increases, while the older, less active stars increase in brightness
when chromospheric activity increases, as is the case for the Sun. The existence 
and relation of photospheric and chromospheric cycles are well established.

On the basis of the variability of the 11-yr solar cycle, which fluctuates between  
approximately 9 and 14 years, and the assumption that stellar activity shares 
similar properties to solar activity, one may presume that
stellar cycles should also show multidecadal
variability. The first attempt to follow {\it changing stellar
activity cycles} was made by Frick et al. (\cite{frick1}), who developed  
and applied a modified wavelet technique suitable for data with gaps, and found a variable cycle 
of one of the targets in the present investigation, HD~100180, from its Ca II index 
record. Subsequent work of Frick et al. (\cite{frick}) and Baliunas et al. 
(\cite{sallie2}) used double-wavelet analysis of the records of stars in the 
Wilson sample to study interdecadal activity variations.

Recent efforts have focused on methods to {\it predict} solar activity based on 
different, earlier observations and on dynamo theory. For critical reviews of 
those methods see Cameron \& Sch\"ussler (\cite{ca_schussler}), and Bushby \& 
Tobias (\cite{bushby}). The knowledge of the cycle pattern of as many active 
stars as possible may yield improved insight on the solar activity seen
in the context of stellar activity. To that end, we have developed a method 
suited for study of multi-decadal variability from gapped datasets. The method 
is presented in detail by Koll\'ath \& Ol\'ah (\cite{zoli_en}, hereafter Paper 
I), where it is applied  to solar data and used to recover a complicated pattern of
multi-scale variability of the Sun in the last few hundred years. In the 
current paper we apply the method to a sample of active stars with photometric 
or spectroscopic records that extend over several decades.

%%%%%%%%%%%%%%%%%%%%%%%%%%%%%
%   O B S E R V A T I O N S
%%%%%%%%%%%%%%%%%%%%%%%%%%%%%

\section{Observations}\label{observ}

\subsection {The sample}

The most important criterion for the selection of stars for analysis was a long record 
with few interruptions. In most of the cases gaps in the records are seasonal interruptions, and we require that the gap does not exceed two seasons. Limited gaps are crucial because
the method of analysis requires equidistant data, and therefore interpolation during gaps in the data (i.e., between observing seasons) is necessary. 

The stellar sample consists of 21 objects, of which the Sun is studied in Paper 
I. Most of the stars have photometric records, and V833 Tau
has a century-long record if augmented with photographic photometry; six stars have
Ca II index records. The photometrically observed active stars are listed in Table~\ref{T1} 
together with their rotational (or orbital) periods, spectral types, v$sin$i, 
and inclination. The last column lists the extent of the records, in 
years.  Single stars are AB~Dor, LQ~Hya, V410~Tau, and FK~Com (about its possible binarity see Kjurkchieva \& Marchev \cite{diana}), and the 
rest (V833~Tau, EI~Eri, V711~Tau, UZ~Lib, UX~Ari, HU~Vir, IL~Hya, XX~Tri, HK~Lac 
and IM~Peg) are synchronised, close binaries. In the case of binaries 
Table~\ref{T1} gives the orbital period and for single stars the rotational 
period. In addition we study six objects from the Wilson sample 
(Wilson, \cite{wilson}): HD~131156A, HD~131156B, HD~100180, HD~201091, HD~201092 
and HD~95735. In our analysis, those targets are considered to be
effectively single stars, though HD~131156A with 
HD~131156B and HD~201091 with HD~201092 form wide, visual binaries. Their rotational 
periods, spectral types and length of the Ca index records are listed in 
Table~\ref{T2}.

%---------------------------------------  Table 1
\begin{table}[tbh]
  \caption{The stellar sample. I. Photometric observations. }\label{T1}
  \begin{center}
  \leavevmode
  \footnotesize
  \begin{tabular}{lrlrrc}
  \hline
  \noalign{\smallskip}
  star & rot. per.$^a$   &  sp. type & \multicolumn{1}{c}{v$sin$i} & \multicolumn{1}{c}{i} & \multicolumn{1}{c}{time-base}\\
       & (days)      &          &\multicolumn{1}{c}{(\kms )} & \multicolumn{1}{c}{(\fok)}  & \multicolumn{1}{c}{(years)} \\
  \noalign{\smallskip}
  \hline
  \noalign{\smallskip}
AB Dor    &  0.515 & K0V       & 91    & 60 & 18 \\
LQ Hya    &  1.601 & K2V       & 27    & 65 & 25 \\
V833 Tau  &  1.788 & K5V       &  6.3  & 20 & 20 \\
 \multicolumn{1}{c}{"}      &  \multicolumn{4}{l}{photographic + photometric 
data}   & 109\\
V410 Tau  &  1.872 & K4        & 74    & 70 & 34 \\
EI Eri    &  1.947 & G5IV      & 51    & 46 & 28 \\
FK Com    &  2.400 & G4III     &155    & 60 & 28 \\
V711 Tau  &  2.838 & G5IV/K1IV & 41    & 40 & 30 \\
UZ Lib    &  4.768 & K0III     & 67    & 50 & 17 \\
UX Ari    &  6.437 & G5V/K0IV  & 39    & 60 & 23 \\
HU Vir    & 10.388 & K1IV/III  & 25    & 65 & 17 \\
IL Hya    & 12.905 & L0III/IV  & 26.5  & 55 & 20 \\
XX Tri    & 23.969 & K0III     & 21    & 60 & 21 \\
HK Lac    & 24.428 & K0III     & 20    & 65 & 50 \\
IM Peg    & 24.649 & K2III     & 26.5  & 70 & 29 \\
  \noalign{\smallskip}
  \hline
  \end{tabular}
  \end{center}
The stellar parameters are from Strassmeier (\cite{klaus_sum}), and the references 
therein.\\
$^a$ in case of synchronized binary the orbital period is given\\
\end{table}

%---------------------------------------  Table 2
\begin{table}[tbh]
  \caption{The stellar sample. II. Ca index measurements.}\label{T2}
  \begin{center}
  \leavevmode
  \footnotesize
  \begin{tabular}{lrlc}
  \hline
  \noalign{\smallskip}
  star & rot. per.   &  sp. type & \multicolumn{1}{c}{time-base} \\
       & (days)      &          & \multicolumn{1}{c}{(years)} \\
  \noalign{\smallskip}
  \hline
  \noalign{\smallskip}
HD 131156A = $\xi$ Boo A &  6.25  & G8V    & 35 \\	
HD 131156B = $\xi$ Boo B & 11.1  & K4Ve   & 35 \\
HD 100180 = 88 Leo & 14.3  & G0V    & 34 \\
HD 201092 = 61 Cyg B & 35.5   & K7V    & 35 \\
HD 201091 = 61 Cyg A & 36.1   & K75    & 35 \\
HD 95735 = GJ 411 & 54.7   & M2V    & 34 \\
  \noalign{\smallskip}
  \hline
  \end{tabular}
  \end{center}
\end{table}

\subsection{Multi-decadal photometry}\label{photdata}
The photometric records contain all published material for each star in the 
sample, in $V$-colour. Those stars are well studied, with numerous papers 
describing the records. Recent summaries for the stars are as 
follows: AB~Dor: J\"arvinen et al. (\cite{jarvinen}), LQ~Hya: Berdyugina et al. 
(\cite{berd1}) and K\H ov\'ari et al. (\cite{zsolt1}), V833~Tau: Ol\'ah et al. 
(\cite{olah_v833}),  V410~Tau: Strassmeier et al. (\cite{klaus_apt}), EI~Eri: 
Strassmeier et al. (\cite{klaus_apt}), FK~Com: Ol\'ah et al. 
(\cite{olah_fkcom}),  V711~Tau: Lanza et al. (\cite{nuccio}) and Strassmeier et 
al. (\cite{klaus_apt}), UZ~Lib: Ol\'ah et al. (\cite{olah_uzlib}), UX~Ari: 
Aarum-Ulv{\aa}s \& Henry (\cite{aarum1}), HU~Vir: Strassmeier et al. 
(\cite{klaus_apt}), IL~Hya: Strassmeier et al. (\cite{klaus_apt}), XX~Tri: 
Strassmeier et al. (\cite{klaus_apt}), HK~Lac: Ol\'ah et al. 
(\cite{olah_hklac}), IM~Peg: Rib\'arik et al. (\cite{gabor}). For several stars 
the records were updated through 2007-2008 using new photometry from the Vienna 
APT (Strassmeier et al. \cite{apt}).

For further details, see the references of the cited papers. Periods of sparsely-sampled data leading to large gaps, sometimes seen in the onset of records, were disregarded.

\subsection{Ca II index measurements}\label{specdata}

The Ca II measurements were collected at the Mount Wilson Observatory, with two sets 
of equipment on two different telescopes, from 1966 to 1978 and after, for 
details see Vaughan et al. (\cite{vaughan}). The relative Ca emission is the 
flux ratio of two 0.1 nm passbands centred on the cores of the H and K lines and two 2 nm passband 
in the nearby continuum.

%%%%%%%%%%%%%%%%%%%%%%%%%%%%%%%
%%%%%%%%  M E T H O D  %%%%%%%%
%%%%%%%%%%%%%%%%%%%%%%%%%%%%%%%

\section{Method}

We apply a time-frequency analysis called short-term Fourier transform (STFT) to 
the photometric and spectroscopic records to study time variability of the 
activity cycles that occur in the sample of twenty stars. As stated
earlier, the analysis requires equidistant data, and therefore we had to interpolate through 
gaps among observations. To interpolate, we used a smoothing spline. 
Before calculating the spline interpolation, 
we removed the rotational signal and the next four strongest components in the 
periodogram near the rotation period; the removal of the next four strongest
signals addresses removal rotation signals that are non-sinusoidal in shape or 
other activity changes on similar time scales from the record. 
Because rotational modulation is subject to change owing to differential 
rotation, removing the rotational signal (and the additional signals of similar 
timescale) was done separately for subsets in the record,
typically around 200 days long, which were then averaged for the spline interpolation. We argue that such a procedure is necessary, because unevenly sampled rotational modulation may alter the average seasonal light level, and consequently cause a false signal in the analysis
for multidecadal variability. This step is especially important for stars with high amplitude 
rotational modulation of large amplitude, arising from, e.g.,  high 
axial inclination to our line of sight. The FWHM of the 
Gaussian used in the time series is usually $\sim$180 days. Note, however, that in the actual calculation we use filtering in the Fourier space instead of temporal filtering. The method is 
explained in detail in Paper I, where tests of the effects of seasonal gaps, 
rotational modulation and observational errors are also found.

The effect of active region growth and decay, as discussed by Messina \& Guinan 
(\cite{messina1}) has little influence on our results because we removed the 
rotational frequencies plus four other signals of similar timescale, thereby 
significantly reducing the amplitudes of the modulation due to rotation, in most 
cases.  Moreover, cycles below about 1.5-2 years are not considered significant, 
as was discussed in Paper I, and below. 

\begin{figure}
\centering
\includegraphics[width=9cm]{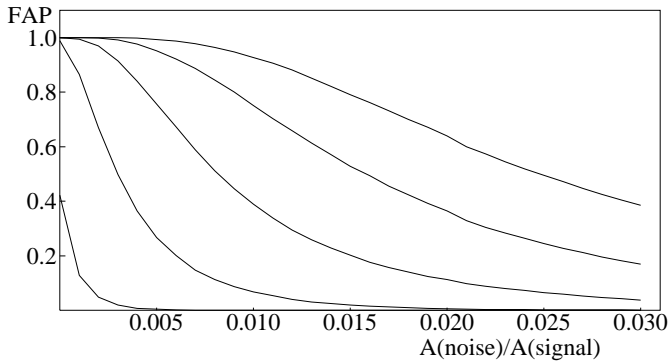}
\caption{False Alarm Probability that a structure appears in STFT when 
Gaussian noise, with different amplitudes, is added to the record of LQ Hya. Five different standard deviations were used and the results are plotted from left to right for 
$\sigma_n$ = 0.01, 0.02, 0.03, 0.04, 0.05 mags. Details are in the text.}
    \label{F0}
\end{figure}

To check the effect of added Gaussian noise on false features in the 
time-frequency diagram, we performed a Monte-Carlo simulation. Here we 
present the results of that test on the record of LQ Hya as an example; data of other stars in 
the sample are of similar quality. First Gaussian noise was added to the 
original observational record. We performed the same preprocessing (averaging 
and spline-smoothing interpolation) as for the standard processing of the 
actual records. The STFT of the difference between
the noisy and the original smoothed data was calculated, and the statistics of 
the maximum amplitude  (largest peak or ridge) was estimated from 10000 cases. 
Fig.~\ref{F0} shows the 
False Alarm Probability $FAP(z)$, e.g. the probability that a structure with 
amplitude larger than $z = A(noise)/A(obs)$ appears in STFT owing to noise
($A(noise)$, and $A(obs)$ is the maximum amplitude in STFT; i.e., $z$ is the 
inverse of the signal-to-noise ratio). The test was performed with different 
standard deviations ($\sigma_n$) of the added Gaussian noise. On the figure from 
left to right the curves belonging to $\sigma_n$ = 0.01, 0.02, 0.03, 0.04, 0.05 
mag are displayed. In case of Gaussian noise
$\sim$0.01-0.02 mag -- which is poorer precision than typical photometric 
precision -- the probability of false signals with amplitudes over 0.01th of the 
highest signal is less than 10\% and fast decreases toward larger astrophysical
amplitudes. 

%%%%%%%%%%%%%%%%%%%%%%%%%%%%%%%
%%%%%%%% R E S U L T S %%%%%%%%
%%%%%%%%%%%%%%%%%%%%%%%%%%%%%%%

%%%%%%%%%%%%% Figure 1 %%%%%%%%%%%%
\begin{figure*}[th!!!!!!!!!!!!!!!!!!!!!!!!!!!!!!!!!!!!!!!!!!!!]
%\begin{figure*}
\centering
\includegraphics[width=5.5cm]{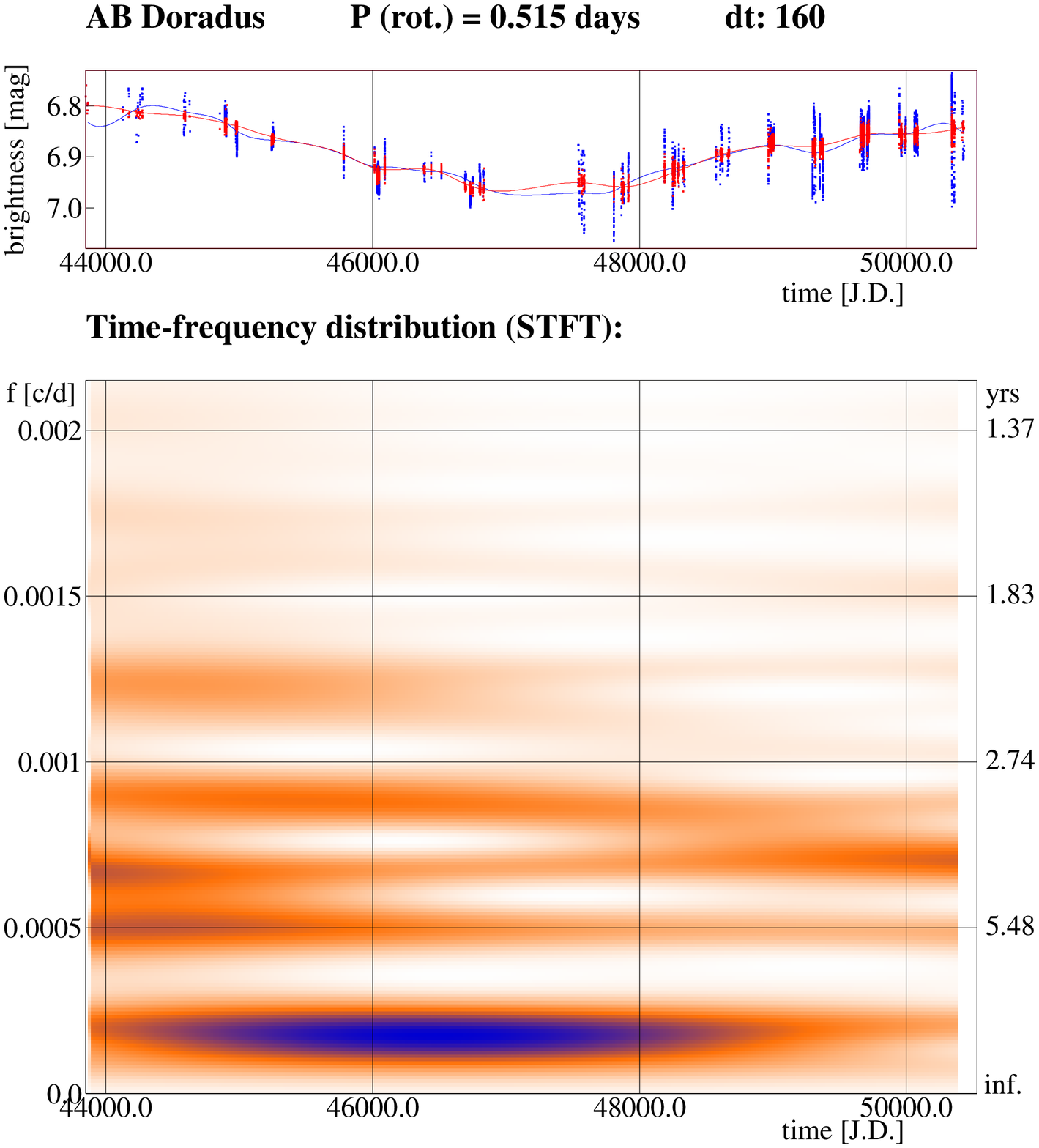}\hspace*{3mm}
\includegraphics[width=5.5cm]{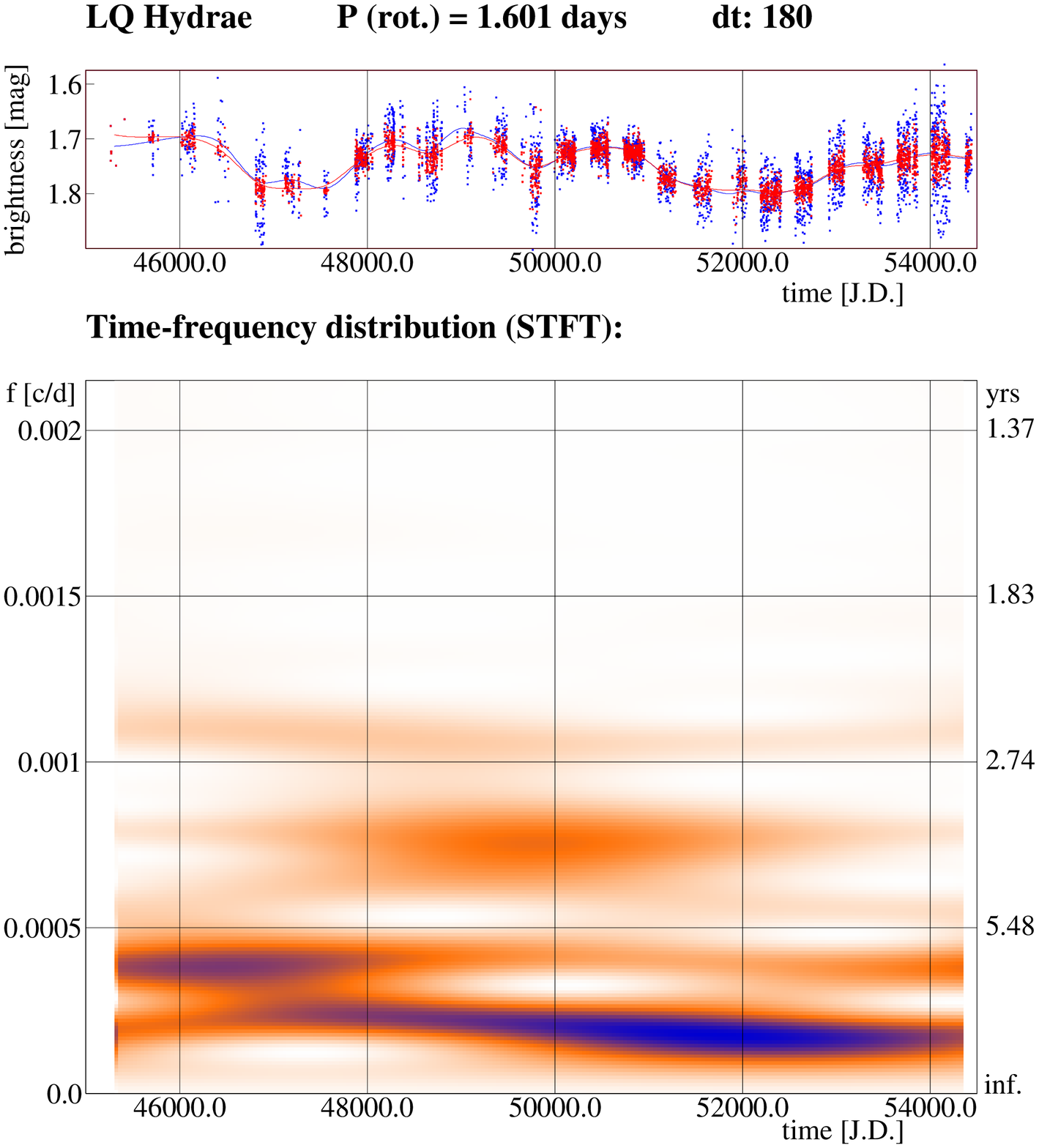}\hspace*{3mm}
\includegraphics[width=5.5cm]{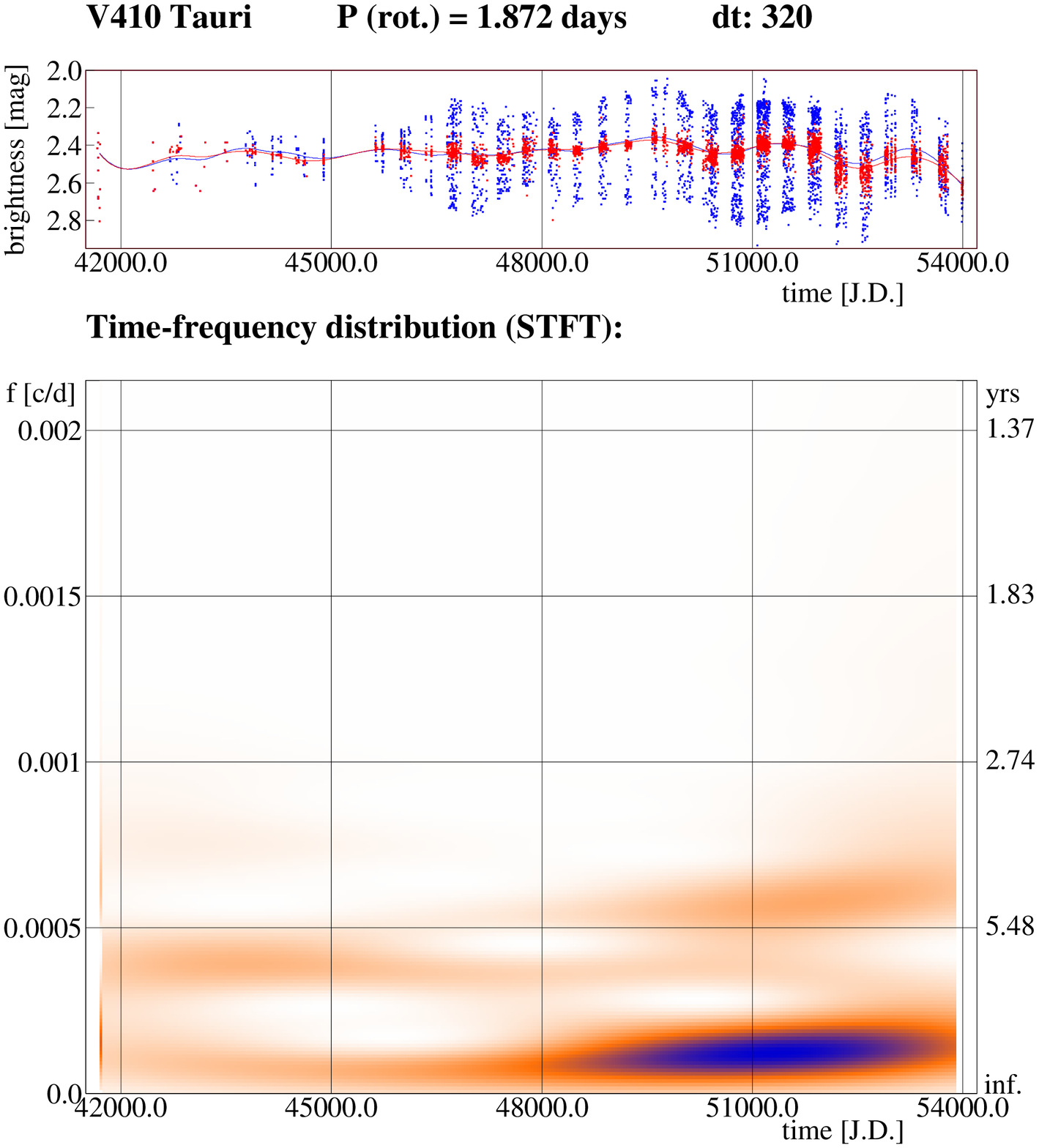}
\vspace*{4mm}
\includegraphics[width=5.5cm]{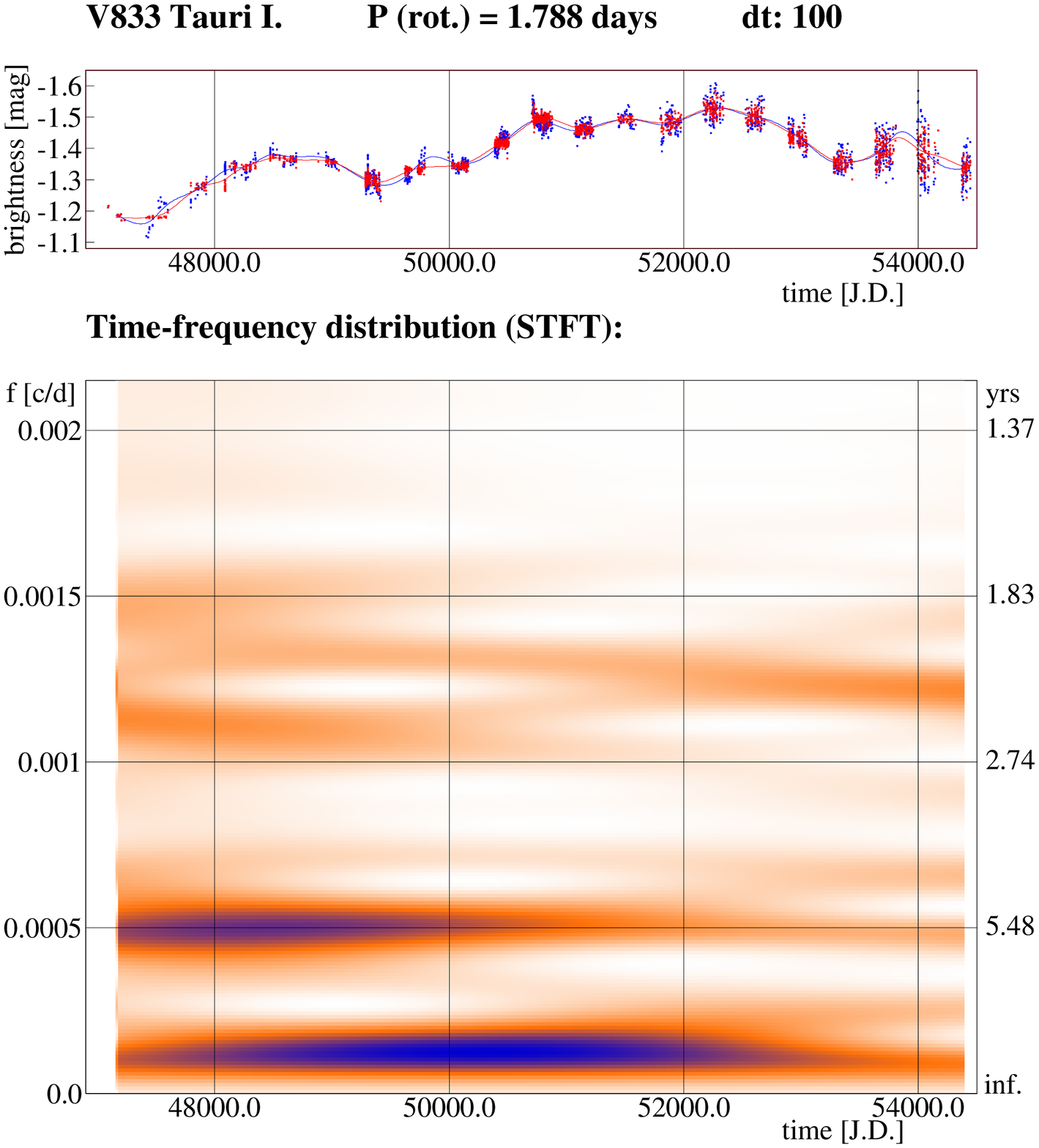}
\hspace*{3mm}\includegraphics[width=5.5cm]{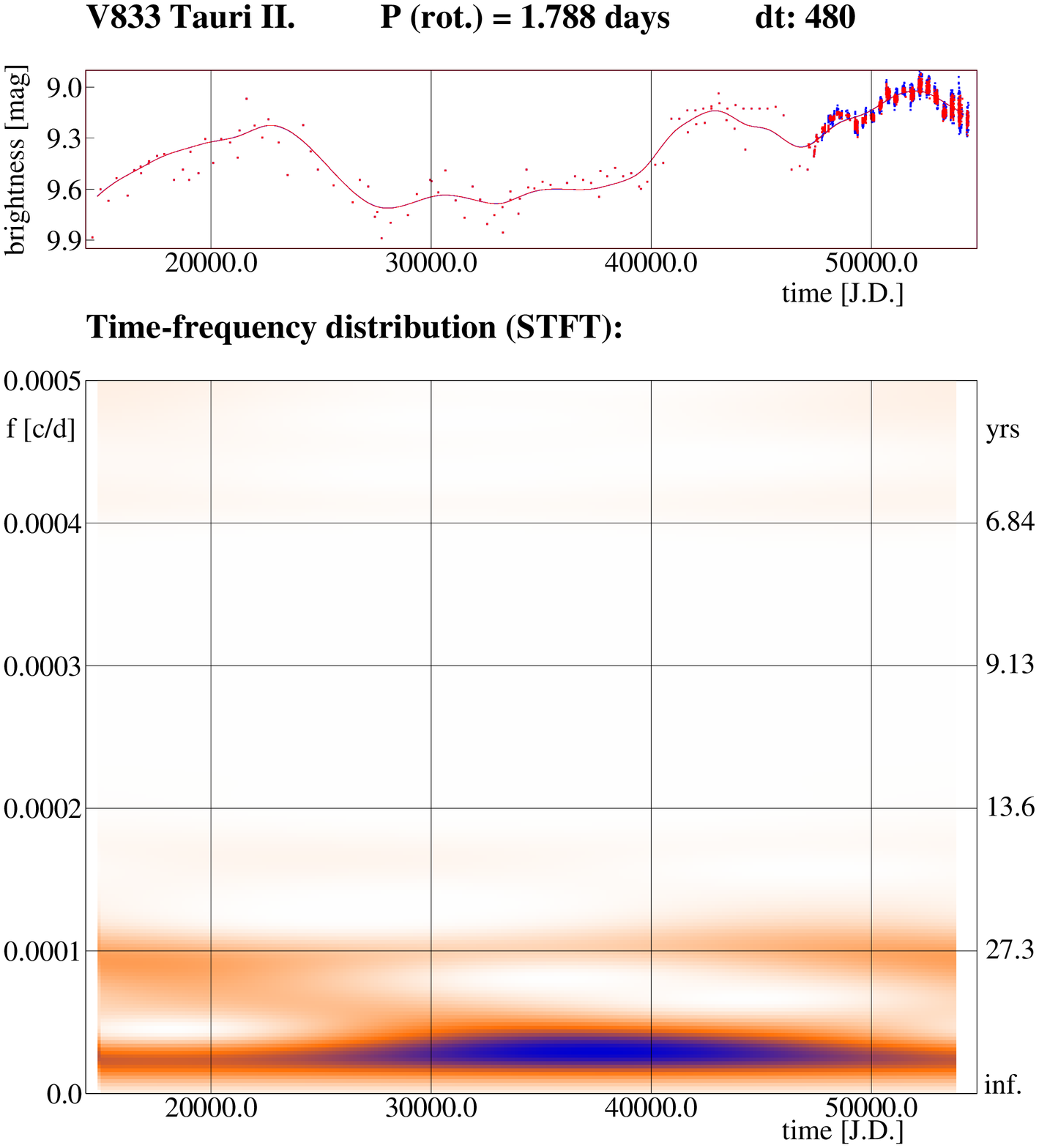}
\hspace*{3mm}\includegraphics[width=5.5cm]{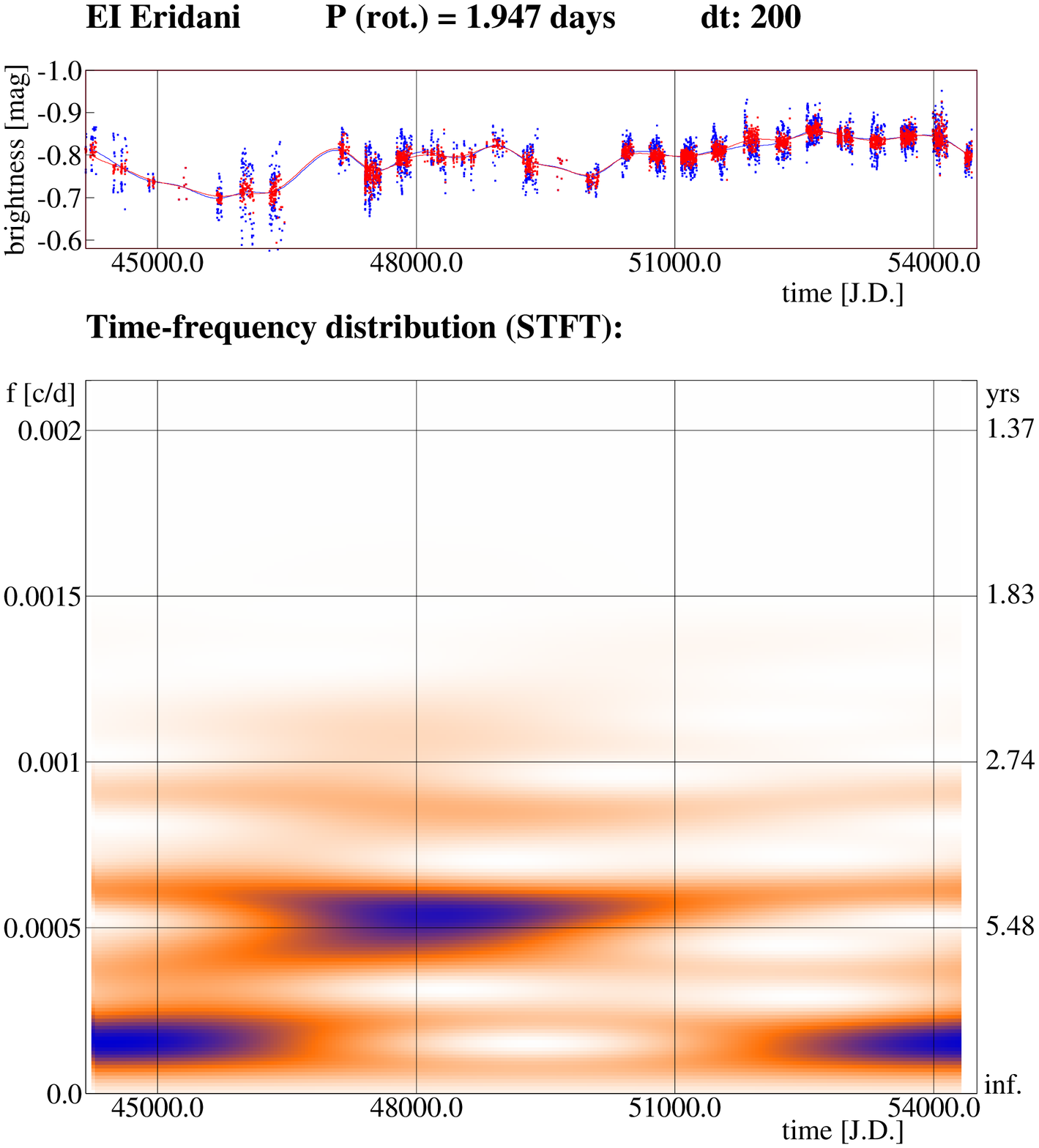}
\vspace*{4mm}
\includegraphics[width=5.5cm]{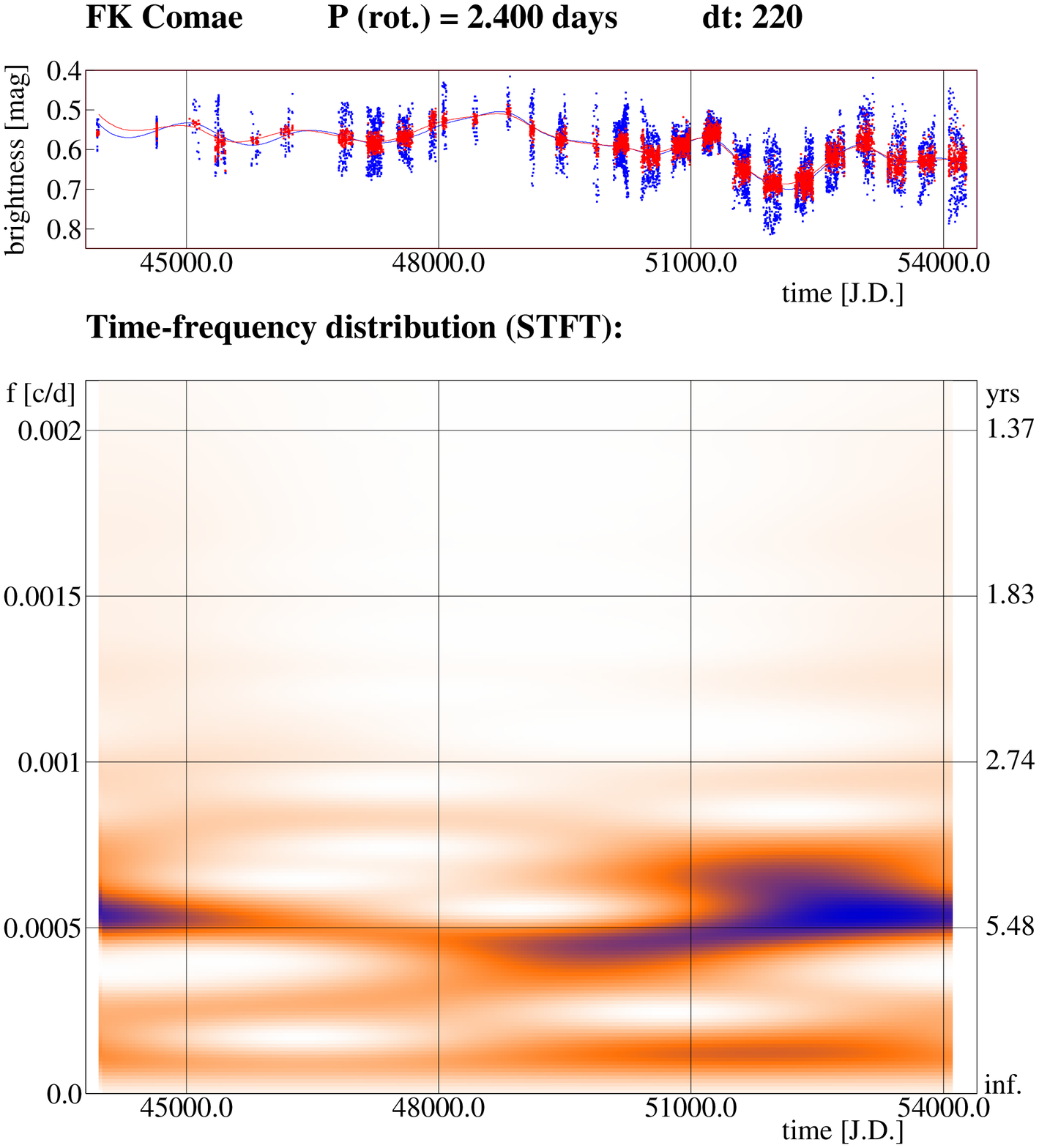}\hspace*{3mm}
\includegraphics[width=5.5cm]{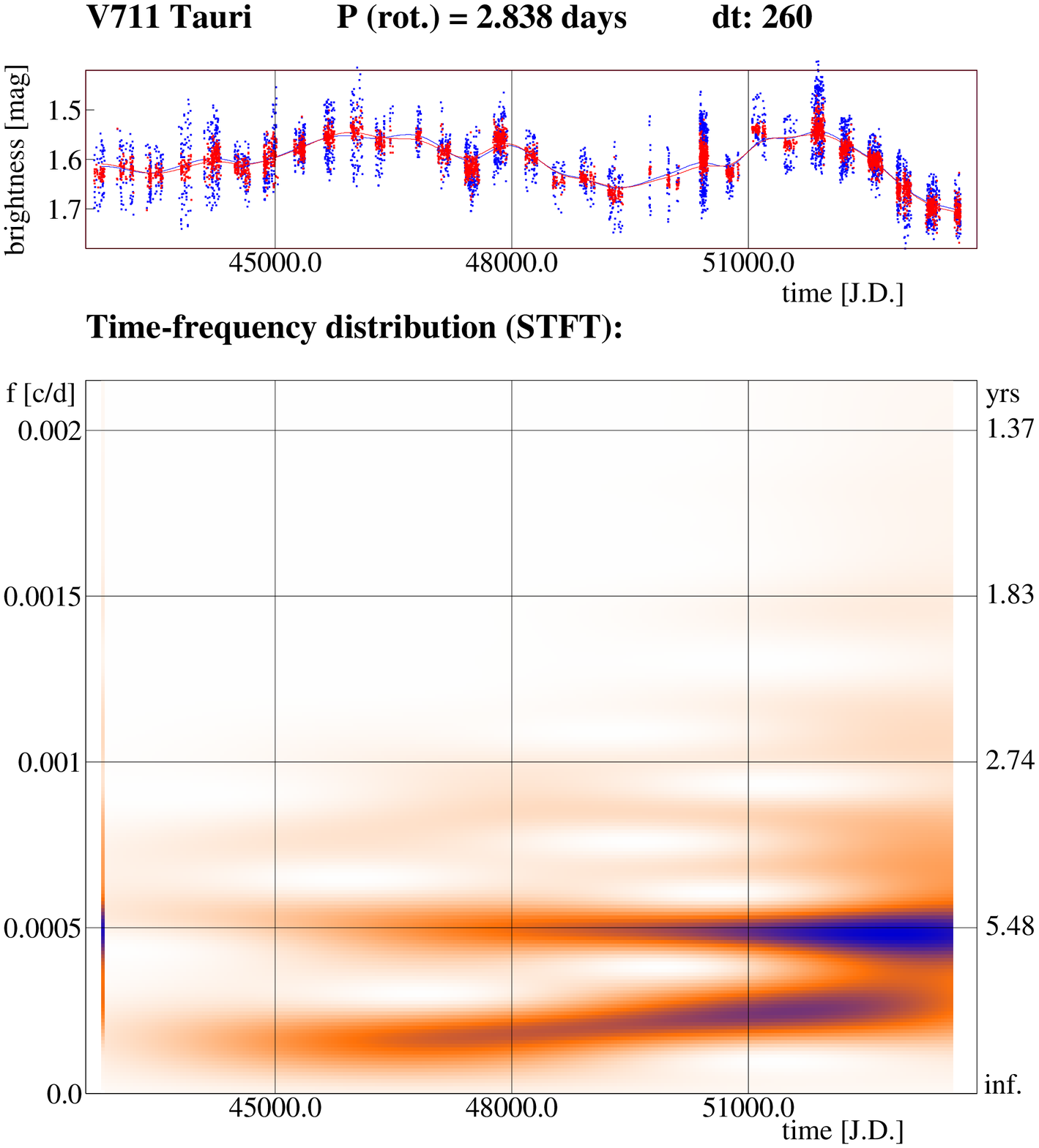}\hspace*{3mm}
\includegraphics[width=5.5cm]{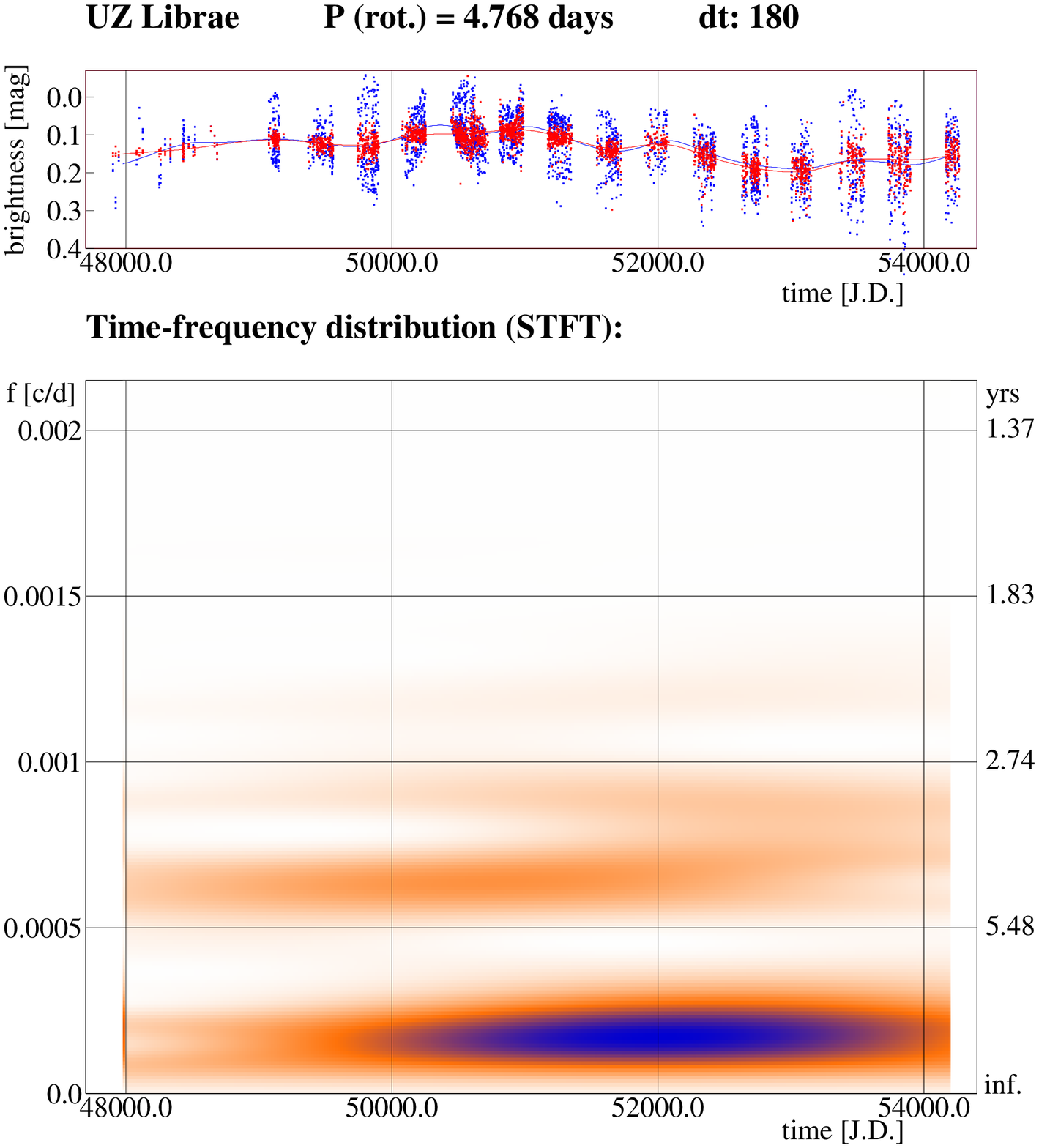}
\caption{Results of the STFT analysis. In the upper panel for each segment 
(i.e., for each star) the observational data (grey, blue in the electronic version), the data prewhitened with 
the mean rotational periods and its five harmonics for each data subset (black, red in the electronic version) 
and the corresponding spline interpolations (solid grey (blue) and black (red) lines) are 
plotted. The lengths of the data subsets are given in the top line of each 
segment. The lower panels show the time-frequency distributions, where darker 
colours denote higher amplitudes. The amplitudes are also modified by the 
amplification factors (see on-line Table~\ref{T3}). Top: AB~Dor, LQ~Hya, 
V410~Tau. Middle: V833~Tau photometric data, V833~Tau photographic+photometric 
data, EI~Eri. Bottom: FK~Com, V711~Tau, UZ~Lib. Details are in the text.}
    \label{F1}
\end{figure*}

%%%%%%%%%%%%% Figure 1 (cont.) %%%%%%%%%%%%

\addtocounter{figure}{-1}
\begin{figure*}[tbh]
\centering
\includegraphics[width=5.5cm]{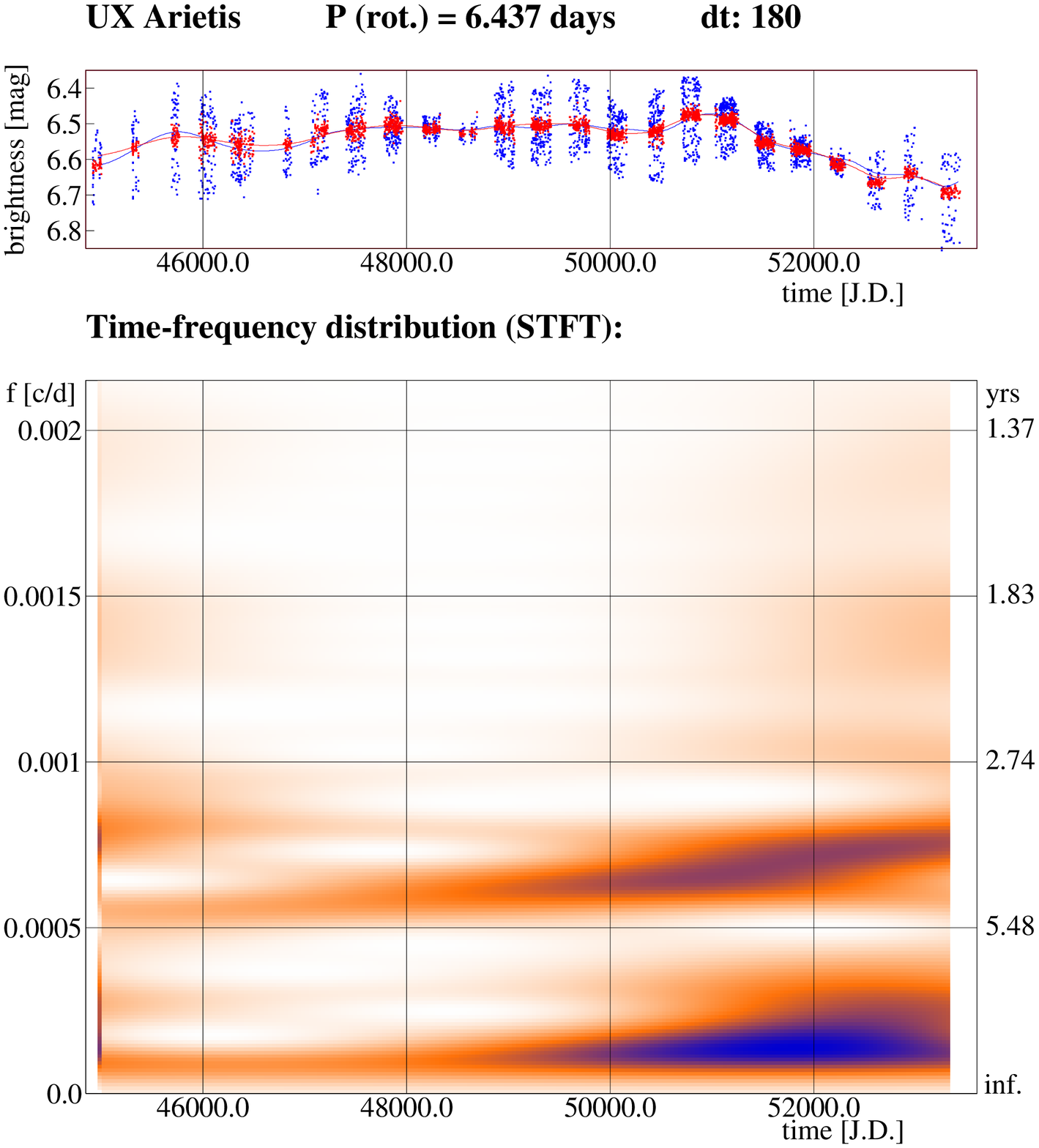}\hspace*{3mm}
\includegraphics[width=5.5cm]{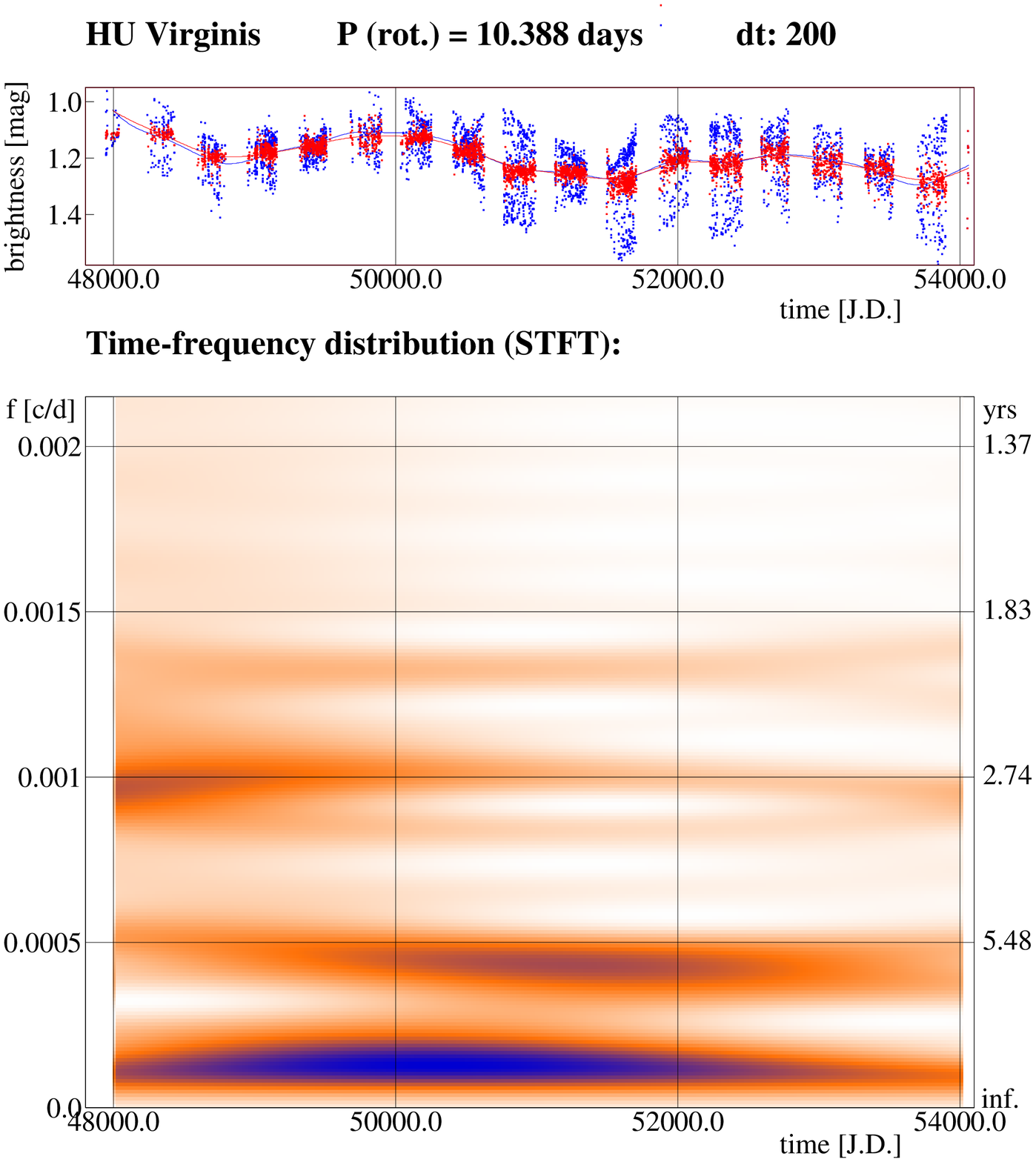}\hspace*{3mm}
\includegraphics[width=5.5cm]{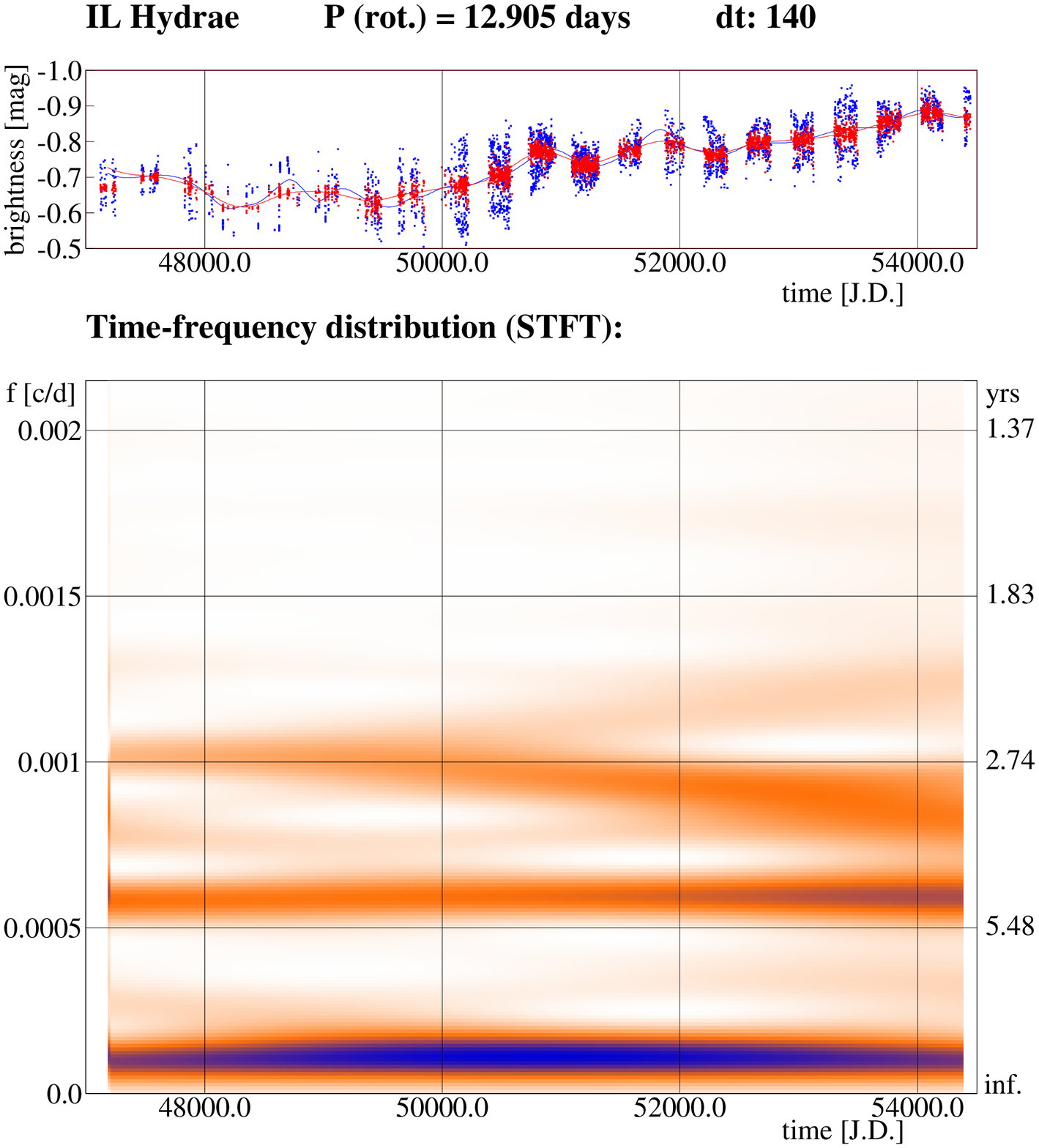}
\vspace*{4mm}
\includegraphics[width=5.5cm]{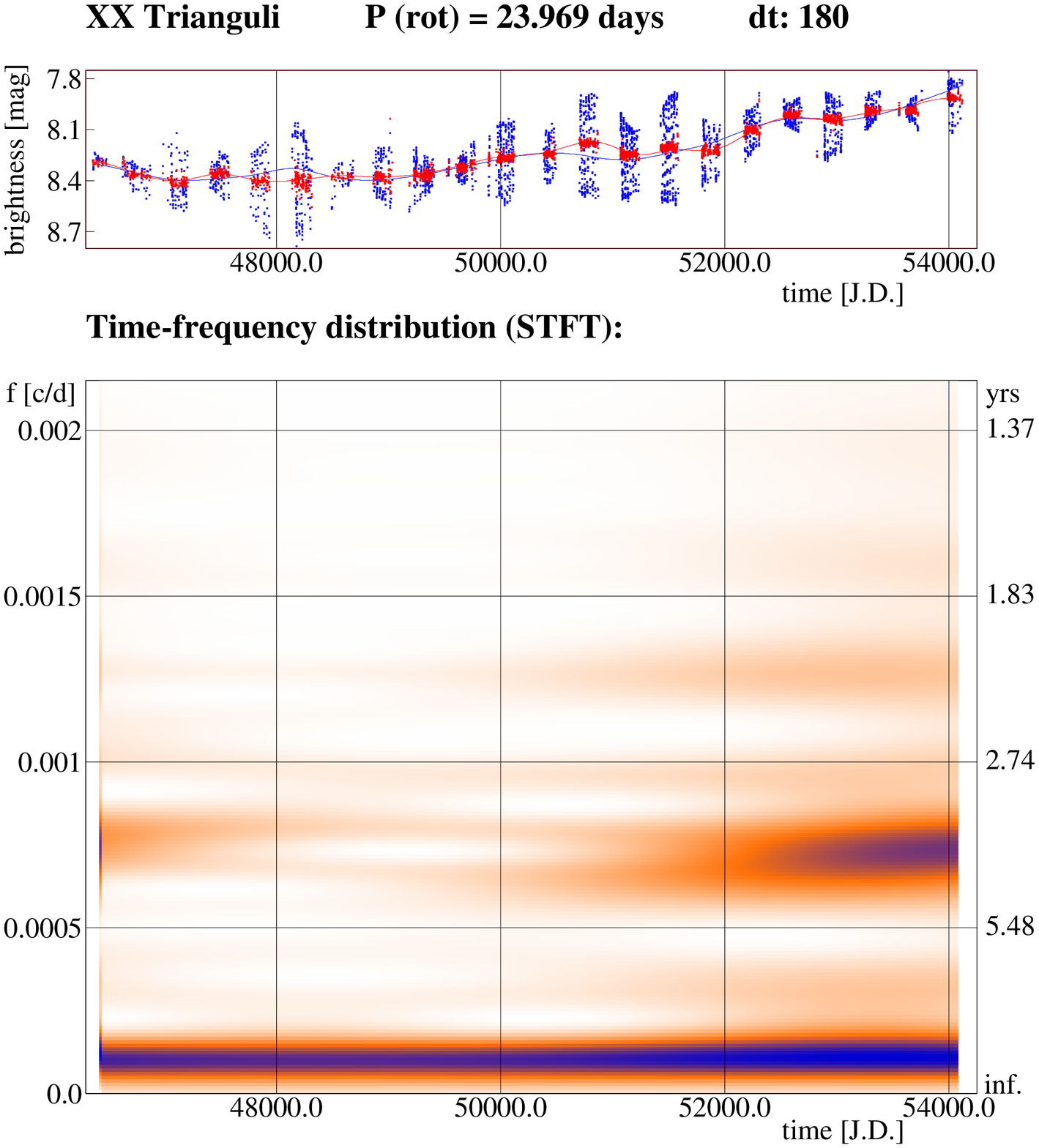}\hspace*{3mm}
\includegraphics[width=5.5cm]{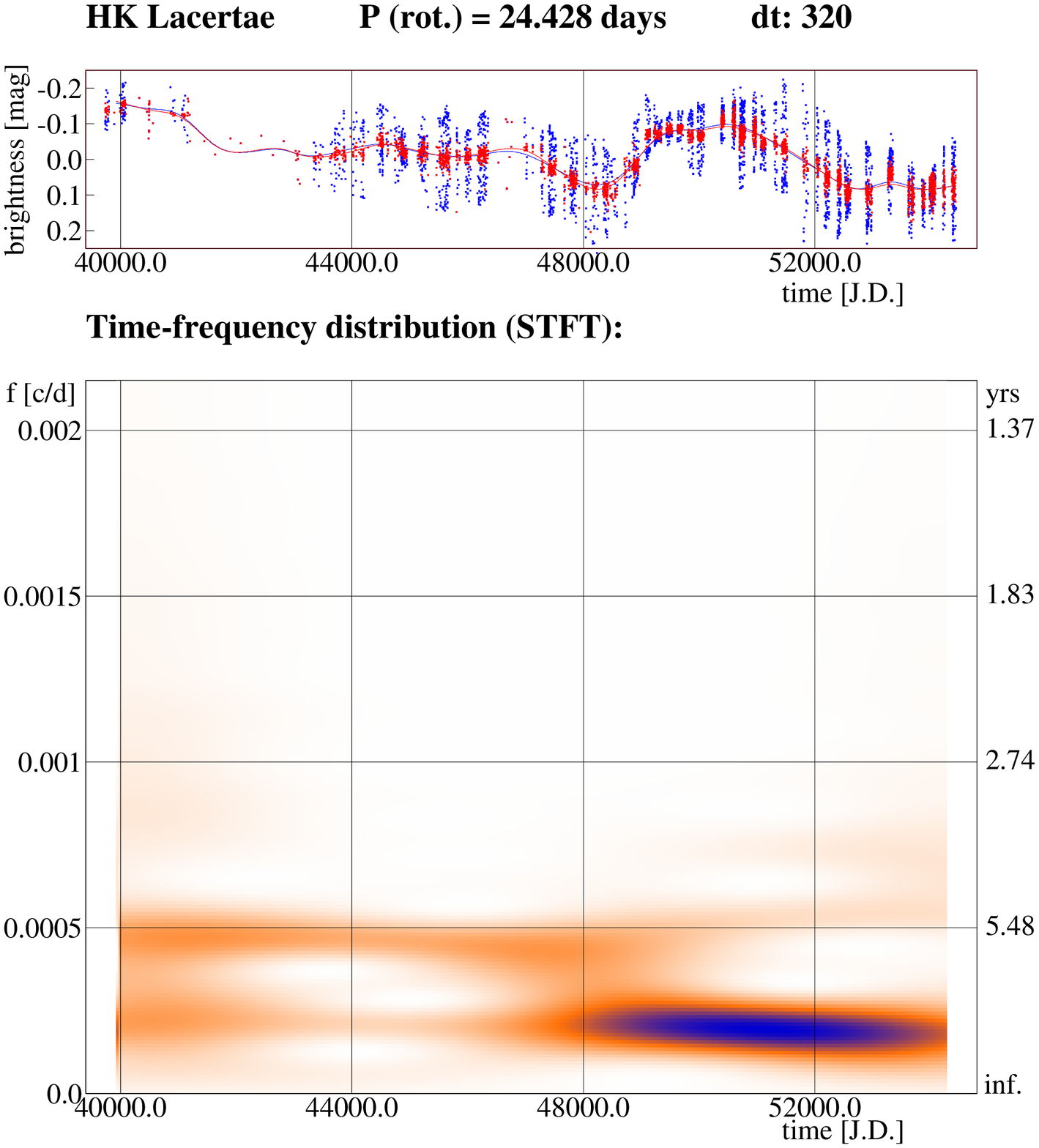}\hspace*{3mm}
\includegraphics[width=5.5cm]{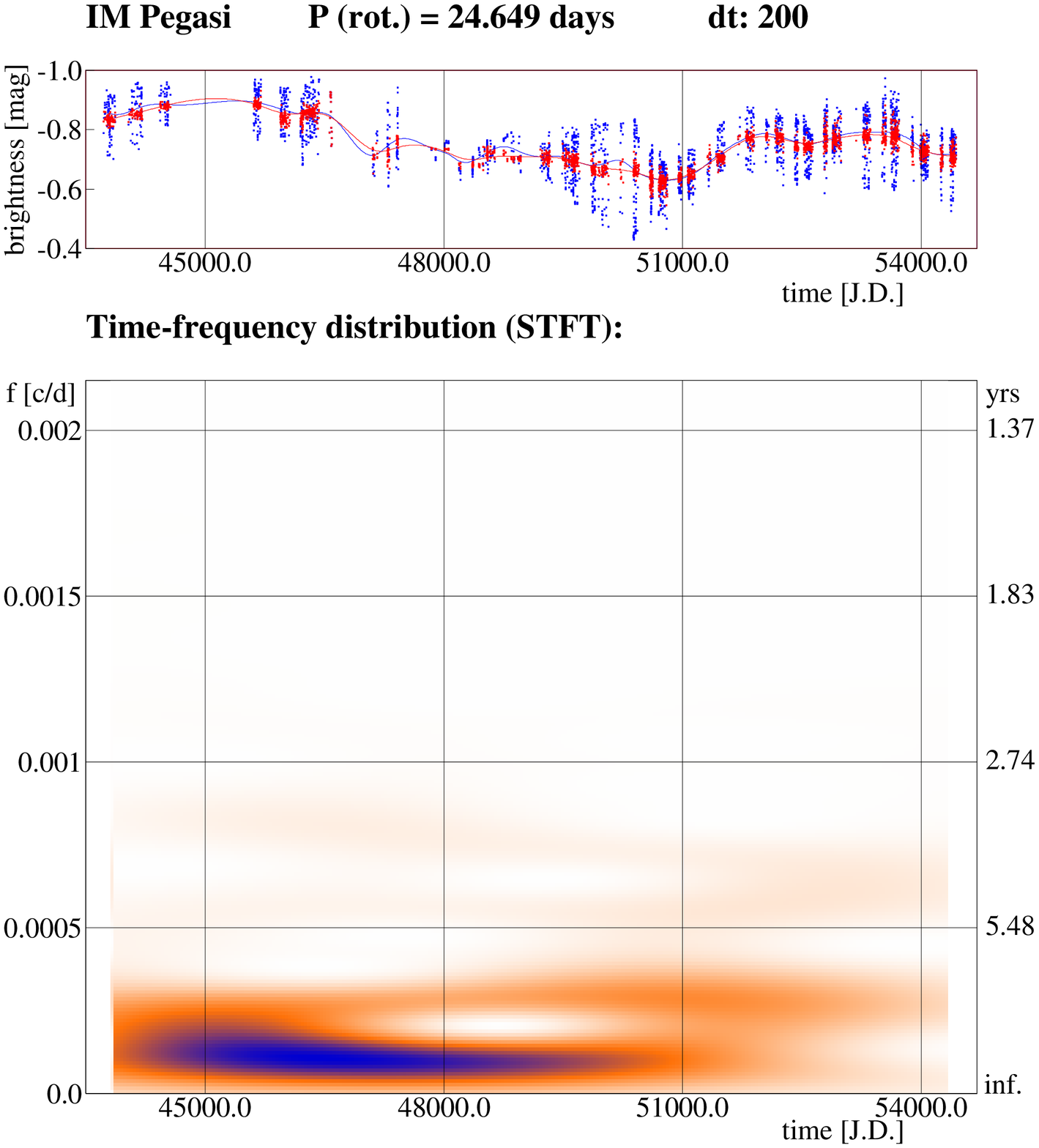}
\caption{{\bf (cont.)} Results of the STFT analysis. Top: UX~Ari, HU~Vir, 
IL~Hya. Bottom: XX~Tri, HK~Lac, IM~Peg. Details are in the text.}
    \label{F2}
\end{figure*}

\section{Results of the time series analysis from photometric data}

From analysis of the records we found changing and multiple cycles for most of the stars. 
Below we present a short description of the cycle pattern for each 
object. The records and STFT results are plotted in Fig.\ref{F1}.

In the time-frequency plots darker colours mean higher amplitudes. The signals 
are modified by different amplification factors that are employed
to make the smaller amplitude signals better visible (see Paper I for more). The amplification factors are given electronically in Table~\ref{T3}. It is not straightforward to estimate the significance of the signals -  moreover, no independent measure can be given. We consider a cycle or cycles 
significant if at least two of these three features are true: a.) the signal runs throughout the period of the observations, b.) it has high amplitude or c.) when two or more components are changing parallel. Cycles shorter than about 1.5-2 years are considered to be insignificant: we 
showed in Paper I that from datasets having yearly gaps such cycles cannot be 
recovered safely. On the other hand, a long-term cycle is not significant if 
its length is commensurable to the time span of the observations, i.e., when the 
data do not cover approximately two full cycles.

%---------------------------------------  Table 3
\onltab{3}{
\begin{table*}[tbh]
  \caption{Amplification factors for the different frequency intervals.}\label{T3}
  \begin{center}
  \leavevmode
  \footnotesize
  \begin{tabular}{lllllll|c}
  \hline
  \noalign{\smallskip}
  star & frequency (1/d) & ampl.  & frequency (1/d) & ampl.  & frequency (1/d) & ampl. & start-end 
J.D.\\
       & interval  & factor & interval  & factor & interval  & factor & 
$-$2400000 d\\
  \noalign{\smallskip}
  \hline
  \noalign{\smallskip}
AB~Dor       & 0.0 - 0.0004 & 0.7 & 0.00041 - & 4.0 & & & 43856-50438\\
LQ~Hya       & 0.0 - 0.0005 & 1.0 & 0.00051 - & 2.5 & & & 45275-54450\\
V833~Tau~I.  & 0.0 - 0.0004 & 1.0 & 0.00041 - 0.001 & 2.0 & 0.0011 - & 5.0 & 
47098-54450\\
V833~Tau~II. & 0.0 - 0.0004 & 0.8 & 0.00041 - 0.001 & 5.0 & 0.0011 - & 9.0 & 
14643-54450\\ 
V410~Tau     & 0.0 -  & 1.0 &  &  &  &  & 41675-54011 \\
EI~Eri       & 0.0 - 0.00035 & 1.0 & 0.00036 -  & 3.5 & & & 44129-54450\\
FK~Com       & 0.0 - 0.0002 & 1.0 & 0.00021 - 0.0006 & 2.5 & 0.00061 - & 4.0 & 
43944-54285\\
V711~Tau     & 0.0 - 0.0004 & 1.0 & 0.00041 - & 3.0 & & & 42719-53703\\
UZ~Lib       & 0.0 - 0.0004 & 1.0 & 0.00041 - & 2.5 & & & 47909-54273\\
UX~Ari       & 0.0 - 0.0005 & 1.0 & 0.00051 -  & 3.5 &  &  & 44928-53440\\
HU~Vir       & 0.0 - 0.0001 & 1.0 & 0.00011 - 0.0008 & 1.0 & 0.00081 - & 3.0 & 
47949-54065\\
IL~Hya       & 0.0 - 0.0005 & 0.8 & 0.00051 - & 3.5 & & & 47129-54450\\
XX~Tri       & 0.0 - 0.0005 & 1.0 & 0.00051 - & 4.0 & & & 46370-54125\\
HK~Lac       & 0.0 - 0.0001 & 0.7 & 0.00011 - 0.00037 & 1.5 & 0.00038 - & 3.2 & 
39751-54450\\
IM~Peg       & 0.0 - 0.0002 & 1.0 & 0.00021 - & 2.0 & & & 43734-54450\\
\hline
HD~131156A   & 0.0 -  & 1.0 &  &  &  & & 39670-52414\\
HD~131156B   & 0.0 - 0.0005 & 1.0 & 0.00051 - & 5.0 & & & 39670-52414\\
HD~100180    & 0.0 -  & 1.0 &  &  &  & & 39898-52418\\
HD~201092    & 0.0 - 0.00014 & 1.5 & 0.00015 - 0.0005 & 1.0 & 0.00051 - & 2.0 & 
39670-52518\\
HD~201091    & 0.0 - 0.0006 & 1.0  & 0.00061 - & 3.0 & & & 39670-52518\\
HD~95735     & 0.0 - 0.0005 & 1.0 & 0.00051 - & 2.0 & & & 39965-52419\\
  \noalign{\smallskip}
  \hline
  \end{tabular}
  \end{center}
The values represent the different amplitudes of the cycles (see text).
\end{table*}
}

{\bf AB~Dor.} The record of AB Dor is not well sampled, the number of 
observations per year is low and gaps longer than two years occur twice. The only clear signal 
is a cycle $\sim$3.3-yr. The other, weaker signal of $\sim$2-yr appears in the 
beginning of the record, and is not considered reliable, and possibly owes to undersampling. 
Variability of high amplitude is reported on the time scale of $\sim$20-yr (Innis et al. \cite{innis}), but the record is not long enough to verify it from our analysis.

{\bf LQ~Hya.} This star is a fast rotating, single star, one of the most 
interesting targets. The light variation is well-sampled by the 
measurements and all gaps are less than one year long. Previously, cycles between 11.4-11.6 years, 6.8-6.5 years and 2.8-3.2 years were determined by Ol\'ah et al. (\cite{cycles1}, 
\cite{cycles2}) from observations spanning a shorter period, and 
more recently, cycles of 13.8$\pm$2.8-yr, and its harmonic 6.9$\pm$0.8 and 3.7$\pm$0.3-yr were found from long term photometry by K\H ov\'ari et al. (\cite {zsolt1}). The present analysis 
recovers two short cycles $\sim$2.5 and 3.6-yr, the latter being stronger in the middle of the dataset.  Another cycle $\sim$7-yr is present, and that period 
increases continuosly to 12.4 years with the same amplitude, 
while a small amplitude $\sim$7-yr signal remains.

{\bf V833~Tau.} The short (20-yr) photometric record of this binary was 
supplemented by archive photographic data (Hartmann et al. \cite{hartmann}), 
thereby extending the dataset to 109 years. We analysed separately the well 
sampled short and more sparse, long records; the long one includes also the photometric 
data. The short photometric record shows two modulations, with the weaker 
$\sim$2.2-yr long and the stronger $\sim$5.2-5.5-yr. Both are present 
throughout the record, which agrees with our previous results 
(Ol\'ah et al. (\cite{cycles1}, \cite{cycles2}). The rotational modulation has 
a small amplitude (few hundredths of magnitude) resulting from the low inclination 
of the star ($\approx20\degr$). This low amplitude of
rotation modulation is an important fact when we study the 
century long dataset from the archive photographic measurements, even though only 1-2 
points/year are available in Hartmann et al. (\cite{hartmann}). However, those 
data represent the stellar brightness, within the usual error of the 
photographic measurements of $\sim$0.1~mag, because the rotational modulation 
itself has a lower amplitude than this value. Thus, a cycle amplitude above $\sim$0.1~ mag 
is well documented by the photographic record, and is considered real. A long-term modulation is 
present during the 109-yr period of observations, with a large amplitude of $\sim$0.9~mag. A long cycle of $\sim$27-30-yr is also recovered.

{\bf V410~Tau.} The only T~Tauri-type star in the sample, V410~Tau shows 
rotational modulation with a large amplitude (often $\sim$0.5-0.7~mag).
However, its mean brightness has remained within $\sim$0.2~mag during the 
34-yr length of the record. The analysis reveals a small-amplitude cycle of $\sim$6.5-6.8-yr, 
which abruptly changes to 5.2-yr near JD 2449500 and afterward slowly decreases in period. 
The long-term trend seems to follow this decrease. Our result supports that of Stelzer et al. 
(\cite{stelzer_et_al}), who found $\sim$5.4-yr 
cycle in yearly mean magnitudes after 1990 ($\sim$2448500).
While Steltzer et al. remarked that the cycle had been out of phase in the earlier observations, we suggest a different, longer, and slowly 
variable cycle length in the beginning of the record.

{\bf EI~Eri.} In the recently available record covering 28 years, two short cycles 
of $\sim$2.9-3.1-yr and $\sim$4.1-4.9-yr are present. They smoothly and 
in parallel increase and decrease; the longer one has a high amplitude between JD 
2446000-50000. A long cycle of $\sim$14-yr with variable amplitude
is also seen. Two cycles of $\sim$2.4 and $\sim$16.2-yr had earlier been 
found by Ol\'ah et al. (\cite{cycles1}). Subsequently, using a longer record, Ol\'ah et al. (\cite{cycles2}) confirmed the $\sim$2.4-yr cycle and revised the longer one to $\sim$12.2-yr. The first determination of the long cycle was 11$\pm$1 years by Strassmeier et al. (\cite{klaus_apt}); all determinations of the long cycle point to an approximate
decadal cycle. Rotational periods of EI~Eri in different seasons has been investigated in detail by Washuettl et al. (\cite{wasi}), and no correlation is found between the seasonal periods (or multiperiods owing to differential rotation of 2-3 active regions at different latitudes for 
most seasons) and the cycle pattern.

{\bf FK~Com.} One dominant cycle is present in the record; it changes smoothly 
between 4.5 and 6.1-yr.  This is consistent with the earlier result by Ol\'ah 
et al. (\cite{olah_fkcom}), who found quasiperiods of about 5.2 and 5.8-yr in 
the longitudes of starspots, based on a record of 18 years, between JD 
2446800-53200 (1987-2004).

{\bf V711~Tau.} The 30-yr record has the best observational coverage among
our sample. The observations themselves show two waves, of $\sim$18 and 9-yr, 
and they are clearly defined in the time-frequency diagram. Apart 
from those two periods, a cycle of $\sim$5.4-yr is present 
throughout most of the record. A short, weak cycle of $\sim$3.3-yr is also apparent from JD 2445000 onwards. Several papers give estimates of the cycles of this well-observed system, and all are in accordance with the present results: Henry et al. (\cite{henry_et_al}) suggests 
a cycle of $\sim$5.5$\pm$0.3-yr and some evidence for a long one of $\sim$16$\pm$1-yr 
from photometry, Vogt et al. (\cite{vogt_et_al}) reported short cycles of 
$\sim$3.0$\pm$0.2 and $\sim$2.7$\pm$0.2-yr for the polar spot area and for the low 
latitude spots from Doppler images. Lanza et al. (\cite{nuccio}) 
using a longer record of photometry found a cycle of $\sim$3-5-yr with variable 
amplitude and a longer cycle, $\sim$19.5$\pm$2.0-yr. Berdyugina \& Henry 
(\cite{berd2}) using an inversion technique on photometric data between 1975-2006 
found $\sim$5.3$\pm$0.1 and 15-16-yr cycles. Taking into account all these 
independent results both from photometry and spectroscopy, the cycle pattern of 
V711~Tau is well documented.

{\bf UZ~Lib.} The cycle length slowly varies from 4.3 to 3.1-yr. Earlier 
Ol\'ah et al. (\cite{cycles2}) had found a cycle length of 4.8-yr. The long-term 
signal near 15-yr is comparable to the length of the dataset itself, and thus 
cannot be established as a cycle.

{\bf UX~Ari.} Its record contains a significant cycle of $\sim$4.6-yr that
deacreases to $\sim$3.6-yr by the end of the record. A long-term signal seems to 
decrease in parallel with the shorter cycle, from $\sim$25 to $\sim$15 years; however,
that longer period timescale is comparable to the length of the record, and is thus not considered significant. On the other hand, the fact that the long cycle changes  
{\it in parallel} with the short one strengthens its reliability. We note that 
Aarum-Ulv{\aa}s \& Henry (\cite{aarum1}) report an activity cycle of 25-yr 
on UX~Ari.

{\bf HU~Vir.} The dominant cycle length of this star is $\sim$5.5-6-yr, in 
agreement with the values found by Ol\'ah et al. (\cite{cycles1}, 
\cite{cycles2}). The shorter-period signals are considered to be insignificant.

{\bf IL~Hya.} A long gap (of approximately 4 years) occurs in the beginning
of the record, which makes the early part of the record unacceptable for our analysis. Hence, we 
considerably shortened the record for study here; compared to the previous studies of Ol\'ah et al. (\cite{cycles1}, \cite{cycles2}, \cite{cycles3}), we are unable the study of the 13-yr  cycle 
found earlier. However,  we confirm the previous result of Ol\'ah et al. (\cite{cycles3}) on the changing nature of the short cycle (from 3.5 to 4.3-yr in that paper). The present result 
shows that a $\sim$4.4-yr cycle persists throughout the 20-yr record, 
and a short cycle, increasing from $\sim$2.7 to $\sim$3 yr also appears.

{\bf XX~Tri.} Apart from a long trend, only a weak $\sim$3.8-yr cycle
is present with significant amplitude only in the second half of the 
record.

{\bf HK~Lac.} The first cycle lengths of this star were estimated by Ol\'ah et al. 
(\cite{cycles1}) as $\sim$6.8 and $\sim$13-yr. Fr\"ohlich et al. (\cite{froehlich}) 
extended the dataset with measurements from Sonneberg Sky-patrol plates, 
and not only confirmed 
the previous results but also found a third cycle with a length of 9.65-yr. The 
origin of the third cycle lies in the changing cycle lengths of HK~Lac, as was 
first communicated by Ol\'ah (\cite{cycles_iau}). The very beginning of the 
record was omitted from the present analysis becasue of scarce 
sampling. The Sonneberg results start well before the photometric measurements, thereby 
lenghtening the record to 50 years; the photographic measurements partly 
overlap the photometry and fill some gaps. The present results, in agreement 
with all previous ones, shows a cycle varying slowly between $\sim$5.4 and 
$\sim$5.9-yr, together with the cycle increasing from $\sim$10.0 to $\sim$13.3-yr.

{\bf IM~Peg.} One cycle of $\sim$9.0-yr is present, in accordance with 
the $\sim$10.1-yr cycle found by Ol\'ah et al. (\cite{cycles2}). A long-term signal 
of 18.9 years is not considered significant, because it is ten years shorter 
than the record.

\section{Results of the time series analysis from Ca II index data}

Studying stellar activity cycles was one of the main aims of the Wilson project 
(e.g. Wilson \cite{wilson}) and from the Ca II index records, fundamental results on decadal variability have been published. Ours is not the first attempt to find 
not only cycle but also rotational periods from that dataset. However, we wished
to compare, using the method here, the results from 
photometry of stellar photospheres with those from Ca II H\&K spectroscopy, which 
reflects the variability of chromospheres.

\begin{figure*}
\centering
\includegraphics[width=4.3cm]{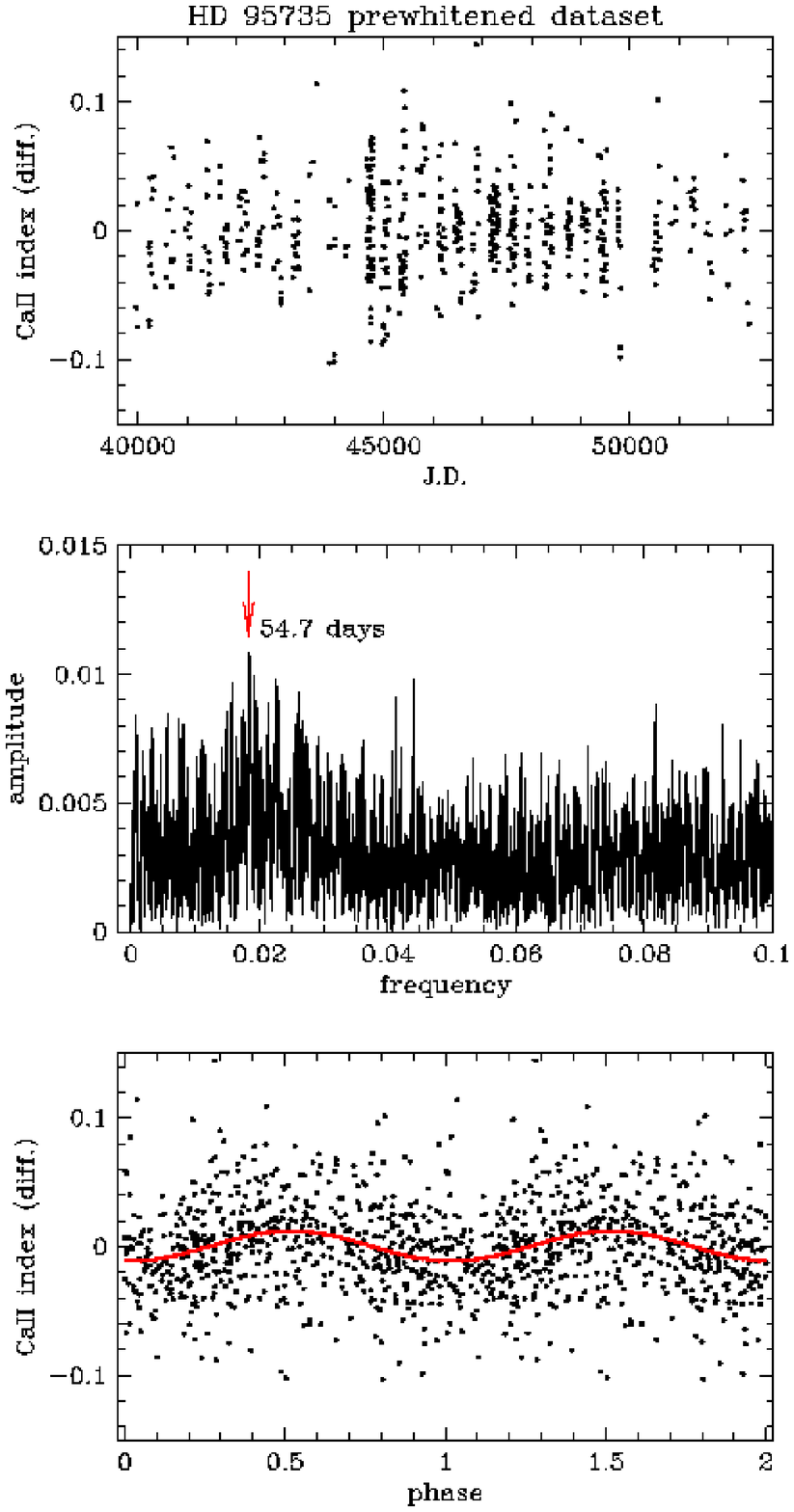}\hspace*{3mm}
\includegraphics[width=4.3cm]{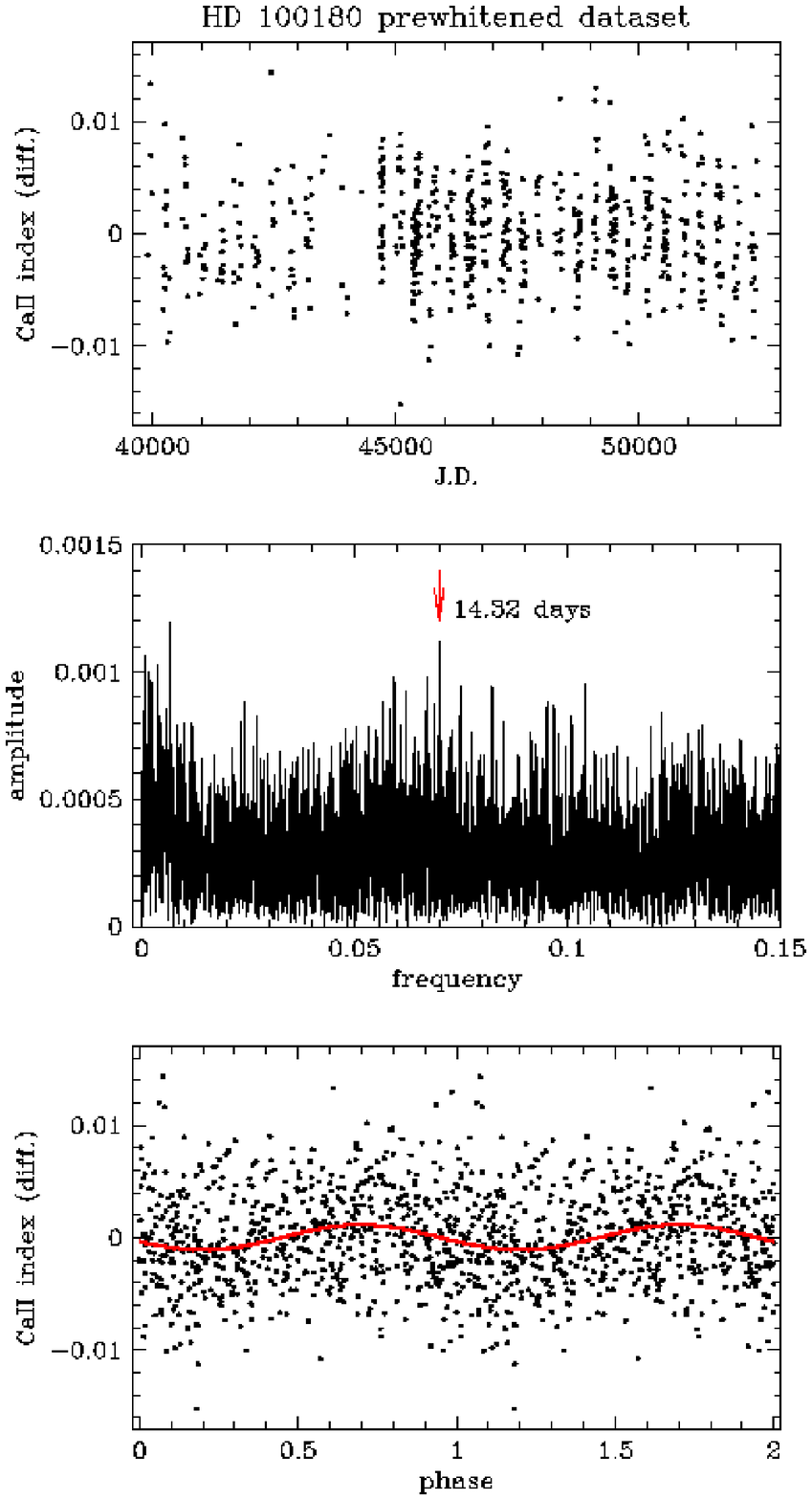}\hspace*{3mm}
\includegraphics[width=4.3cm]{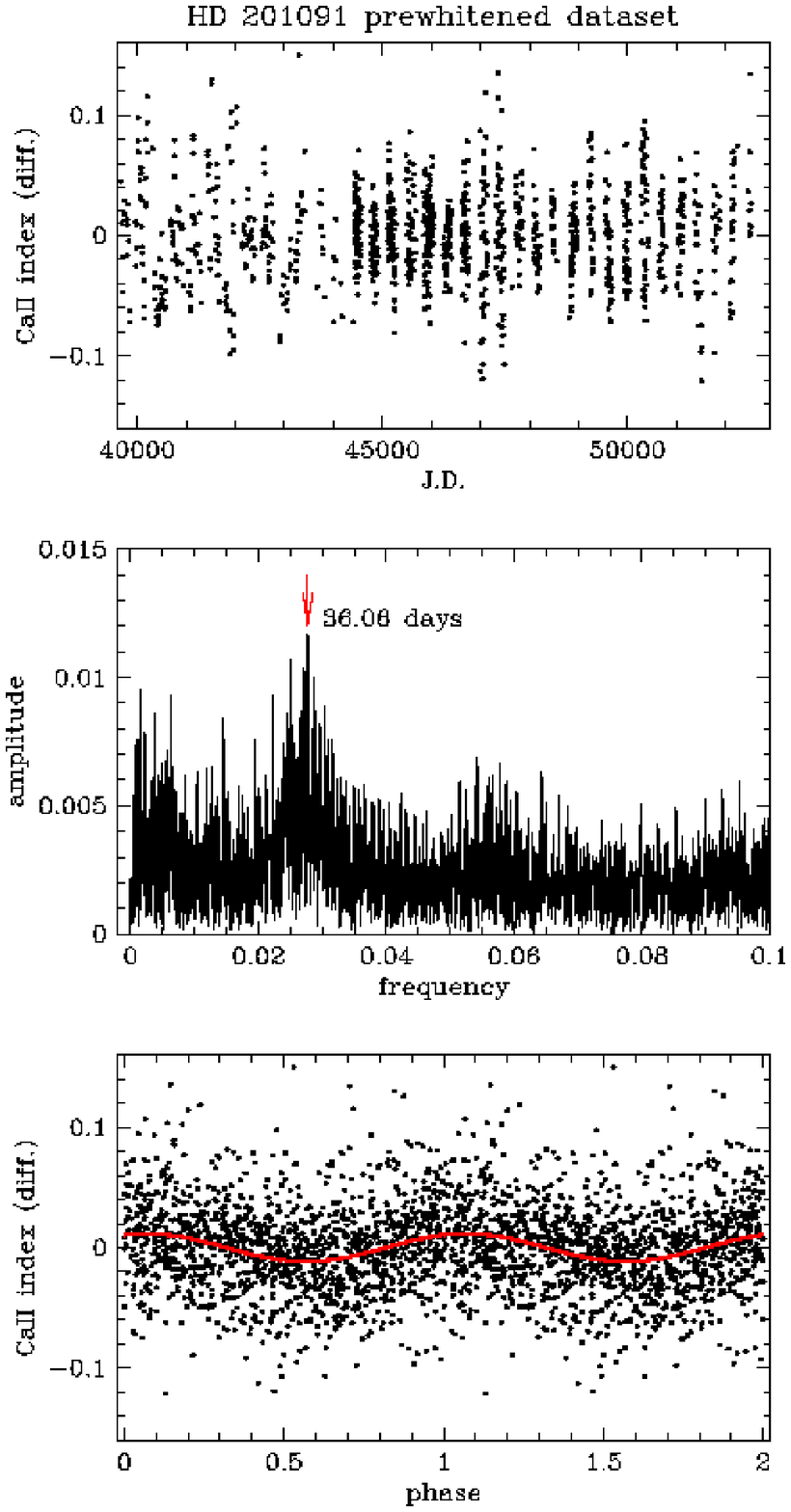}\hspace*{3mm}
\includegraphics[width=4.3cm]{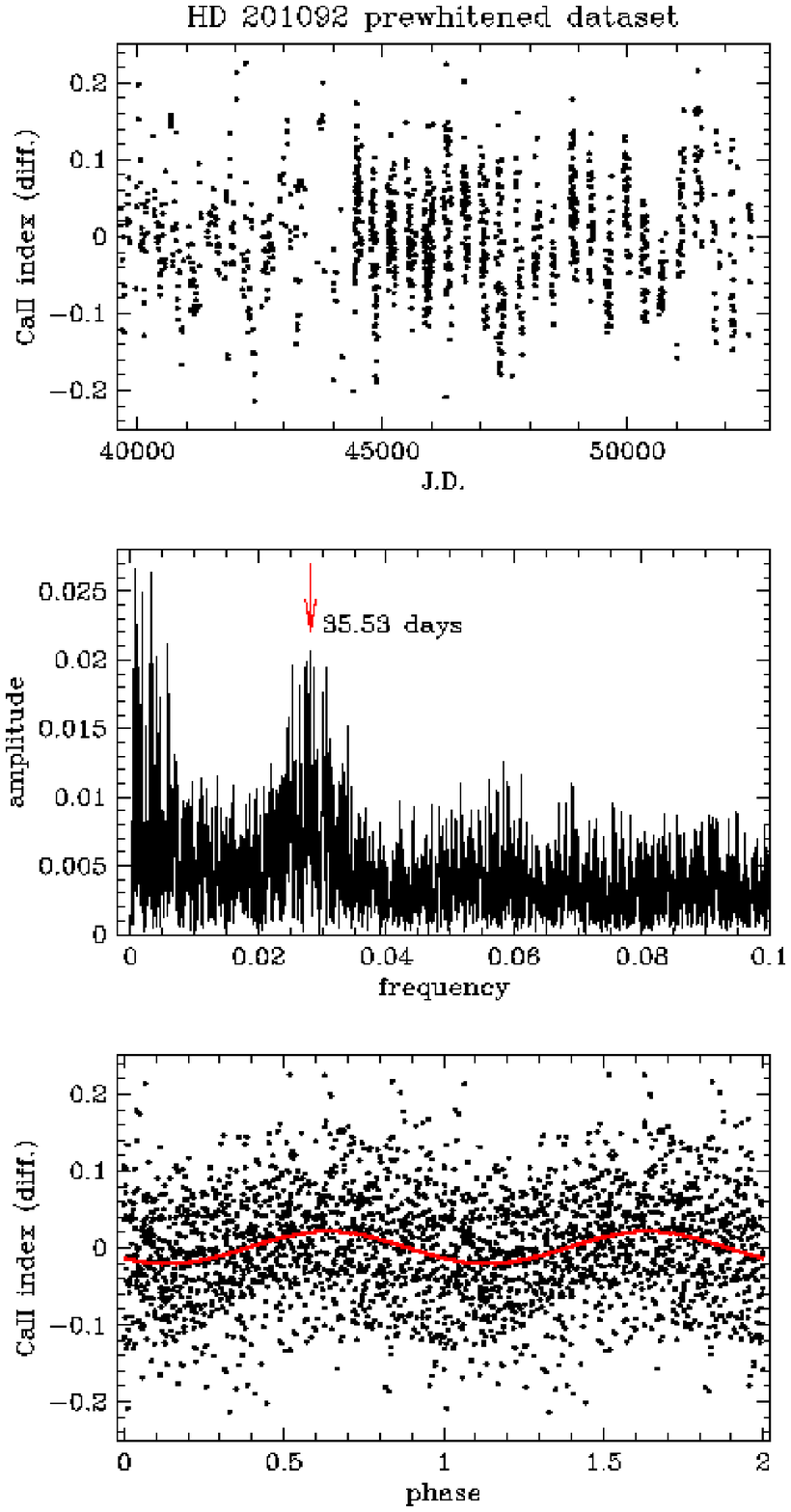}
\vspace*{4mm}
\includegraphics[width=4.3cm]{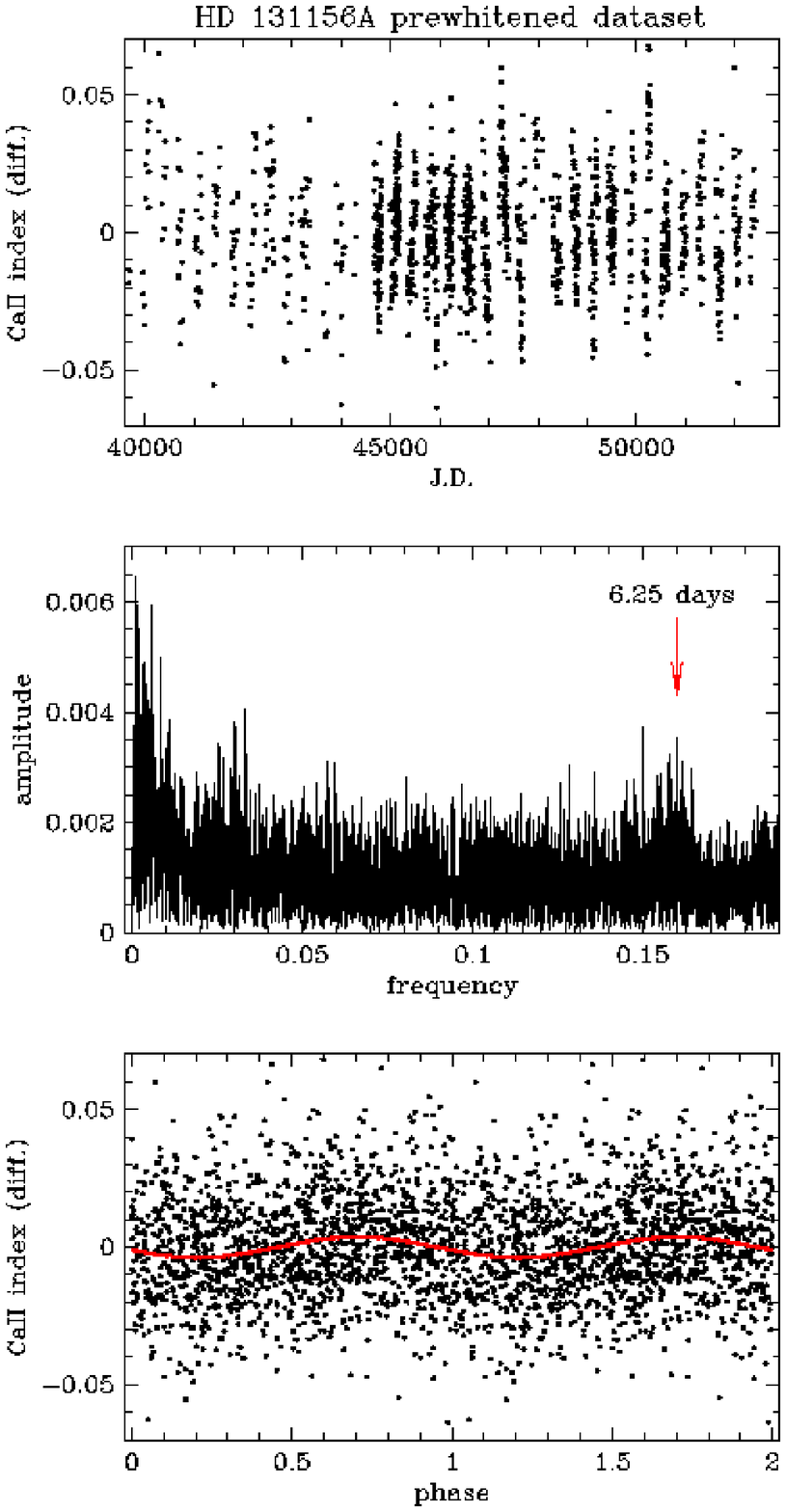}\hspace*{3mm}
\includegraphics[width=4.3cm]{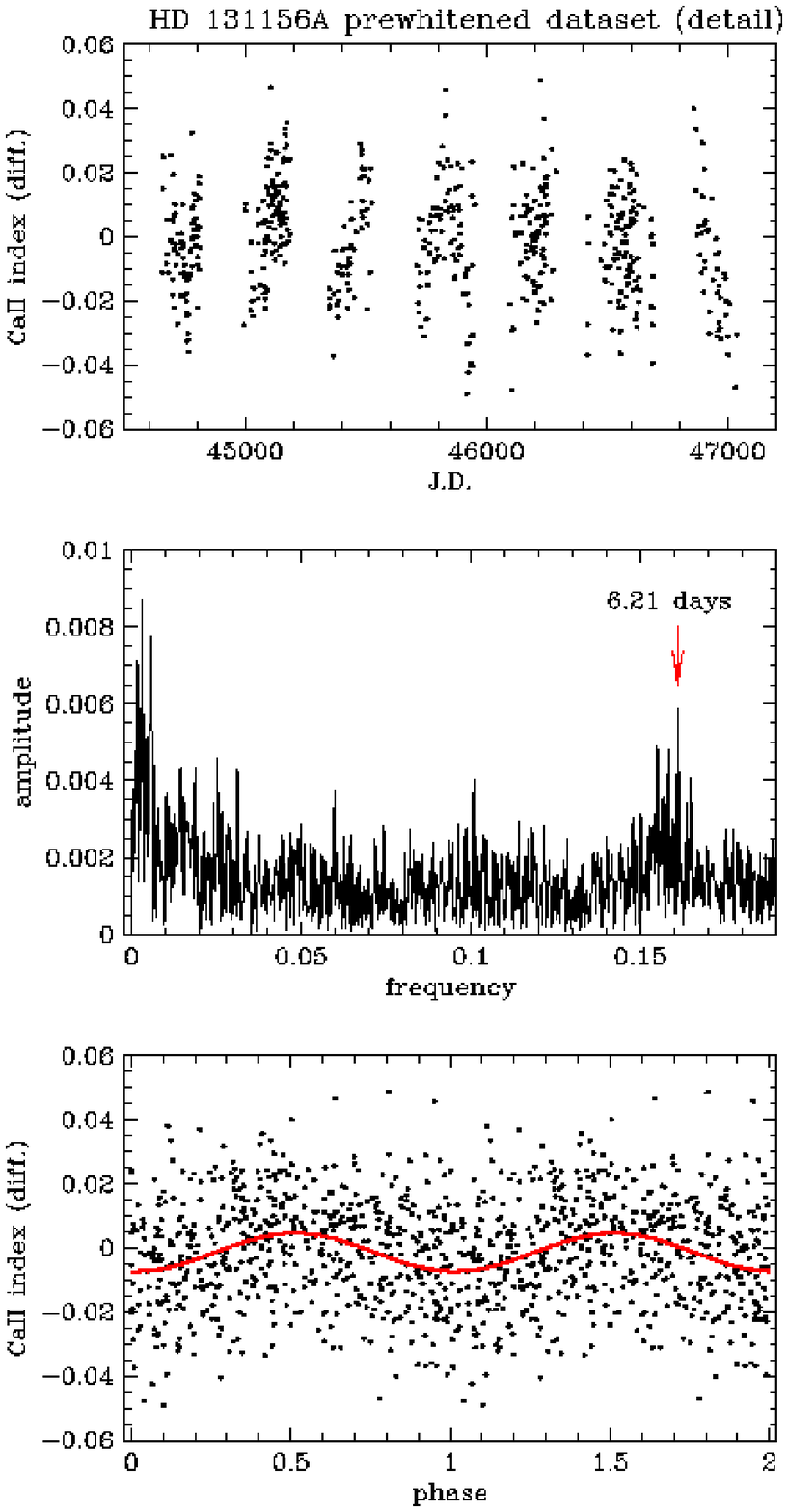}\hspace*{3mm}
\includegraphics[width=4.3cm]{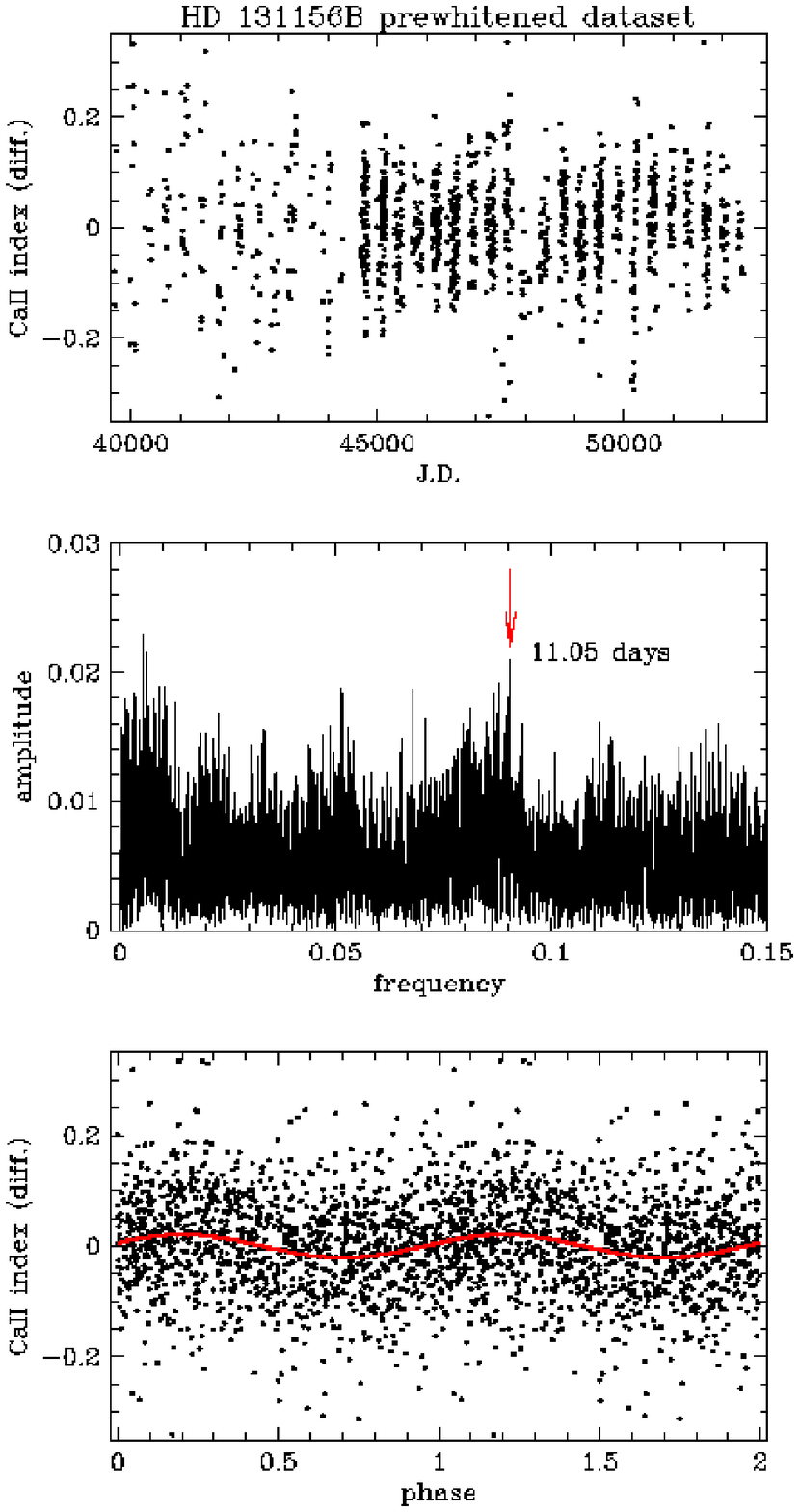}\hspace*{3mm}
\includegraphics[width=4.3cm]{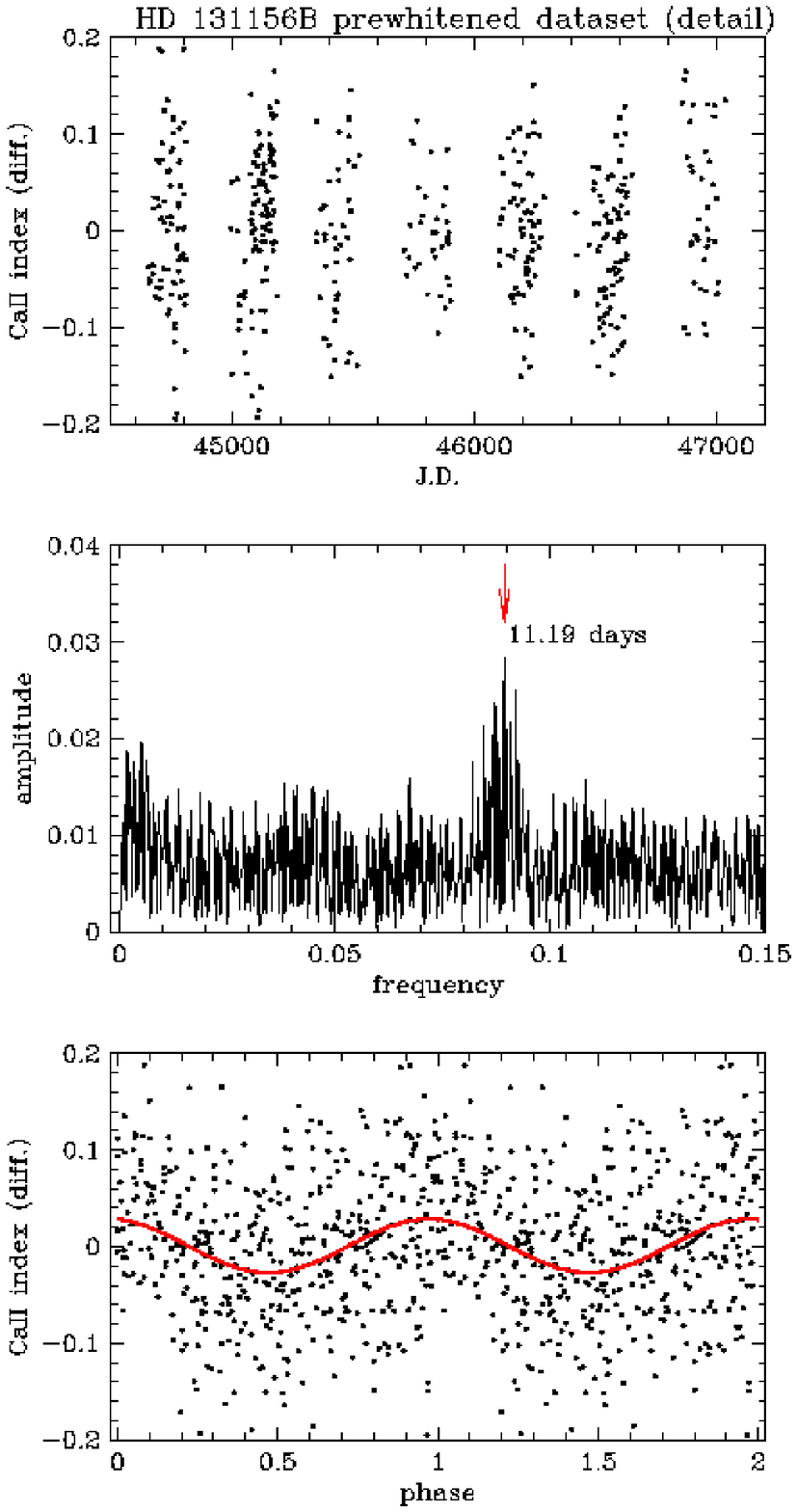}
\caption{Rotational periods from Ca II index data. Upper panel, from left to right: 
HD~95735, HD~100180, HD~201091, HD~201092. Lower panel, from left to right: 
HD~1131156A and detail, HD~131156B and detail. Each segment consists of three 
panels: top: data prewhitened with decadal-scale variations, middle: power 
spectra with the rotational period marked, bottom: data folded on the derived 
period.}
    \label{F3}
\end{figure*}

\subsection{Rotational periods}

The first thorough analysis of rotational modulation from the Ca II index data 
was carried out by Baliunas et al. (\cite{sallie1}). That work is based on records of 112 
stars, among them are all the six objects of the present investigation. All rotational 
periods found by Baliunas et al. (\cite{sallie1}) agree well with the present 
results, which is not always the case in later studies of
other stars in the sample. Fig.~\ref{F3} shows the period analysis of the six stellar records, all of which were prewhitened with the long-term variability (trends and cycle periods). The 
results are presented in Table~\ref{T4} together with periods found in the 
literature. In all cases typical patterns arising from Fourier analysis of 
non-stable signals are seen: e.g., differential rotation of the stars produces 
significant power at frequencies around the main rotational frequency of the amplitude spectra.

We split the records into three nearly equal parts in time and repeated the 
period search. The results are different among the three sets, primarily because of the effects of 
differential rotation, but short lived structures appearing and disappearing at different positions may modify the derived periods as well. The values of periods detected in the subsets of the records are given in the electronic Table~\ref{T5}. Taken into account these results plus those from the literature we estimate uncertainties in the mean derived rotational periods of at least a few tenths of a day in case of shorter periods, and about 1.5-2 days for stars with rotational periods of a few weeks. Seasonal periods determined by Donahue et al. 
(\cite{donahue}) show a larger range of periods than seen in our
analysis because of a multi-seasonal average in our analysis of 
several seasons combined as one subset. Baliunas et al. (\cite{sallie1}) and 
Donahue et al. (\cite{donahue}) used datasets 10 years shorter, and Frick et al. 
(\cite{frick}) and Baliunas et al. (\cite{sallie2}) the same length records as we use in the 
current paper. In Fig.~\ref{F3} (lower panel) we show two examples, HD~131156A and HD~131156B in which the amplitude spectra of the entire record and part of it yield different periods. Note that in the presence of strong differential rotation but without knowing the rotational profile and the inclination of the star, it is nearly impossible to derive a precise mean rotational period.

%---------------------------------------  Table 4
\begin{table}
  \caption{Rotational periods from Ca II index records.}\label{T4}
  \begin{center}
  \leavevmode
  \footnotesize
  \begin{tabular}{lrlrrr}
  \hline
  \noalign{\smallskip}
  Star & \multicolumn{5}{c}{rotation periods in days} \\
 & \multicolumn{1}{c}{(1)}  &  \multicolumn{1}{c}{(2)} & \multicolumn{1}{c}{(3)} & \multicolumn{1}{c}{(4)} & \multicolumn{1}{c}{(5)}\\
  \noalign{\smallskip}
  \hline
  \noalign{\smallskip}
HD 131156A & 6.25  & 6  &  6.31 & -- & 8.2 \\	
HD 131156B & 11.1  & 11 & 11.94 & -- & -- \\
HD 100180  & 14.3 & 14 & --    & 38 & -- \\
HD 201092  & 35.5  & 35 & 37.84 & 36 & 45, 51 \\
HD 201091  & 36.1  & 38 & 35.37 & 36 & 44, 47.5\\
HD 95735   & 54.7   & 53 & --    & -- & 28.5\\
  \noalign{\smallskip}
  \hline
  \end{tabular}
  \end{center}
(1) present paper, (2) Baliunas et al. (\cite{sallie1}), (3) Donahue et al. 
(\cite{donahue}), mean periods, (4) Frick et al. (\cite{frick}, (5) Baliunas et 
al. (\cite{sallie2})
\end{table}

%---------------------------------------  Table 5
\onltab{5}{
\begin{table*}
  \caption{Rotational periods from Ca II index measurements, for the full datasets 
and for three subintervals.}\label{T5}
  \begin{center}
  \leavevmode
  \footnotesize
  \begin{tabular}{lr|rrr}
  \hline
  \noalign{\smallskip}
  \multicolumn{1}{c}{Star} & \multicolumn{1}{c|}{all data}  & 
\multicolumn{1}{c}{1. part} & \multicolumn{1}{c}{2. part} & 
\multicolumn{1}{c}{3. part} \\
       & \multicolumn{1}{c|}{(days)} & \multicolumn{1}{c}{(days)} & 
\multicolumn{1}{c}{(days)} & \multicolumn{1}{c}{(days)} \\
  \noalign{\smallskip}
  \hline
  \noalign{\smallskip}
HD 131156A & 6.25   &  \multicolumn{1}{c}{--} & 6.21  & 
6.15 \\	
HD 131156B & 11.1 &  \multicolumn{1}{c}{--}  & 11.2 & 
11.05 \\
HD 100180  & 14.3   &  \multicolumn{1}{c}{--}  & 14.9  & 
\multicolumn{1}{c}{--} \\
HD 201092  & 35.5    &  30.2  & 36.5 & 
35.53\\
HD 201091  & 36.1   &  37.3 & 36.0 & 
38.4 \\
HD 95735   & 54.7    &  \multicolumn{1}{c}{--}  &  54.5  & 
\multicolumn{1}{c}{--} \\
  \noalign{\smallskip}
  \hline
  \end{tabular}
  \end{center}
Values outside 3$\sigma$ in different subsets 
indicate differential rotation.
\end{table*}
}

\subsection{Time-frequency results}

Below we present the new results of the cycles of each star individually, and 
compare them with the previous results.

\begin{figure*}
\centering
\includegraphics[width=5.5cm]{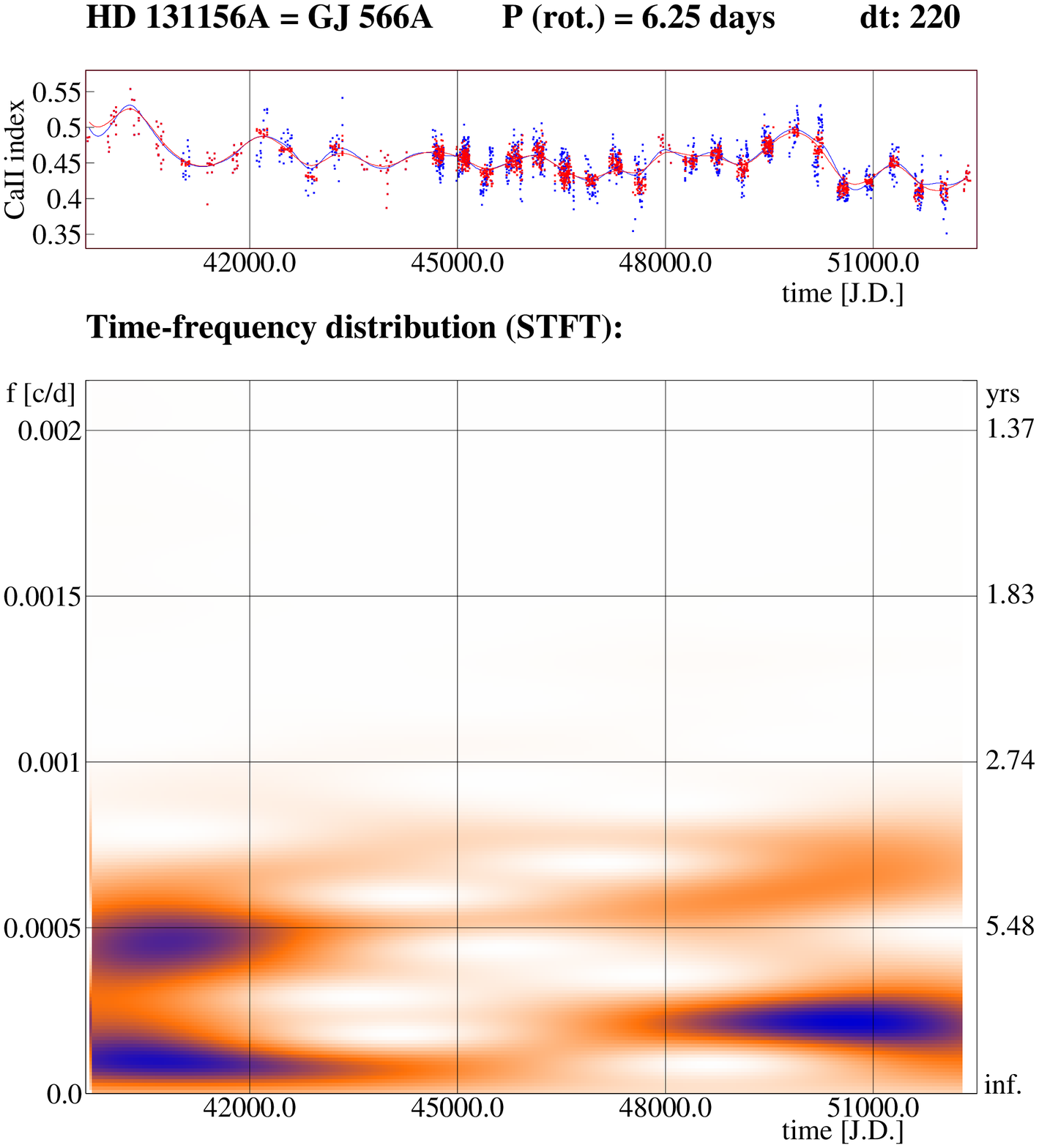}\hspace*{3mm}
\includegraphics[width=5.5cm]{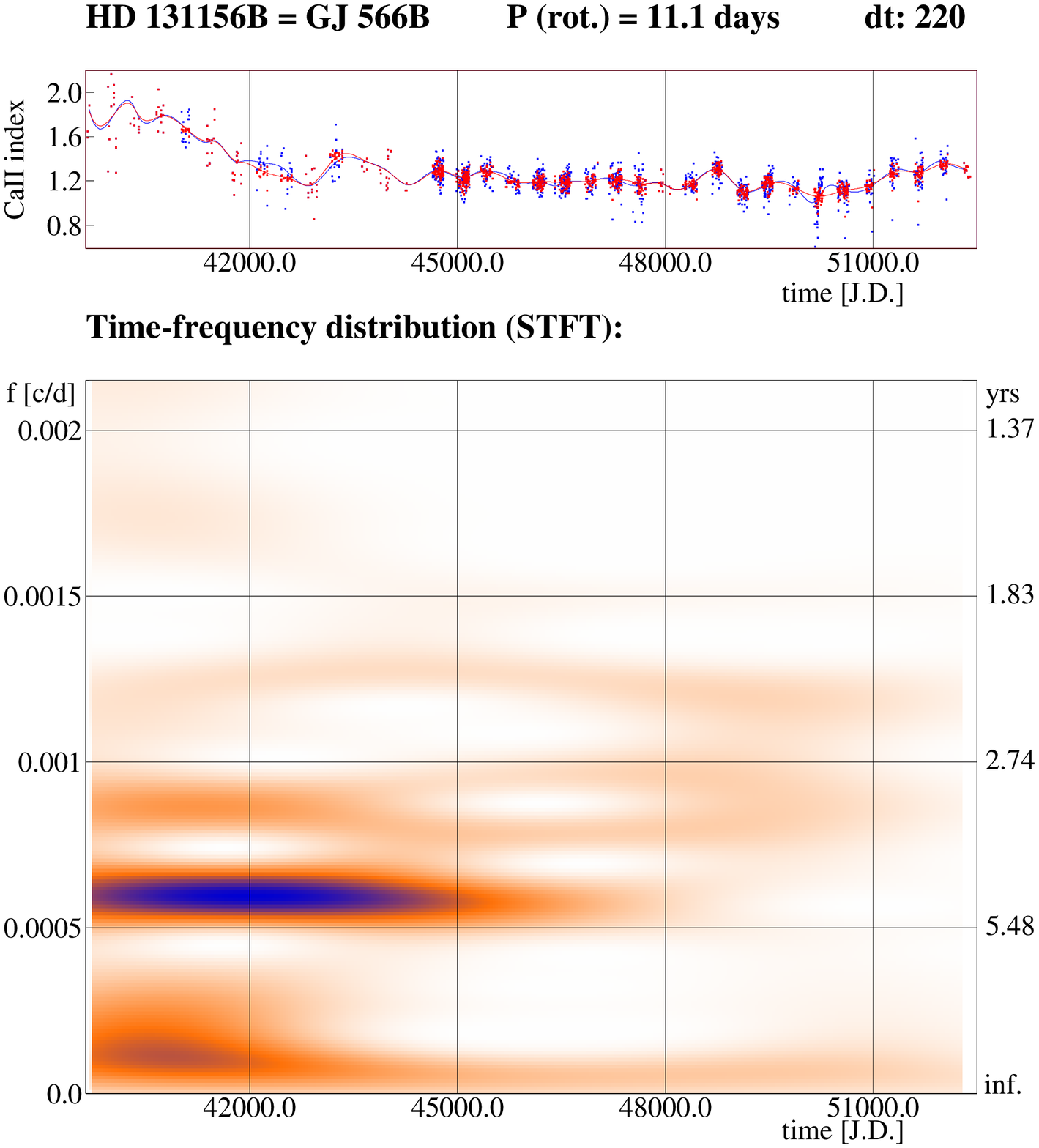}\hspace*{3mm}
\includegraphics[width=5.5cm]{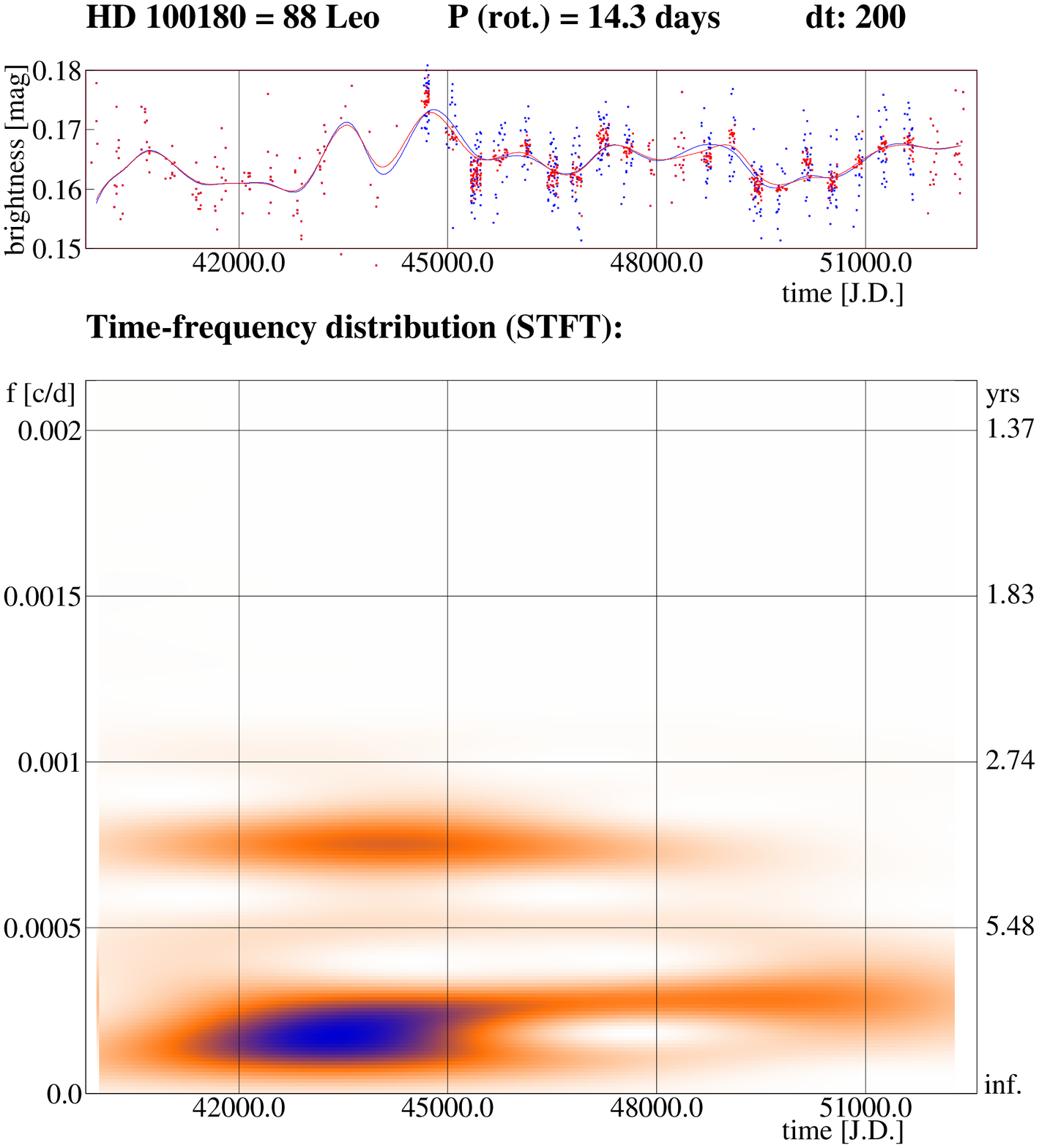}
\vspace*{4mm}
\includegraphics[width=5.5cm]{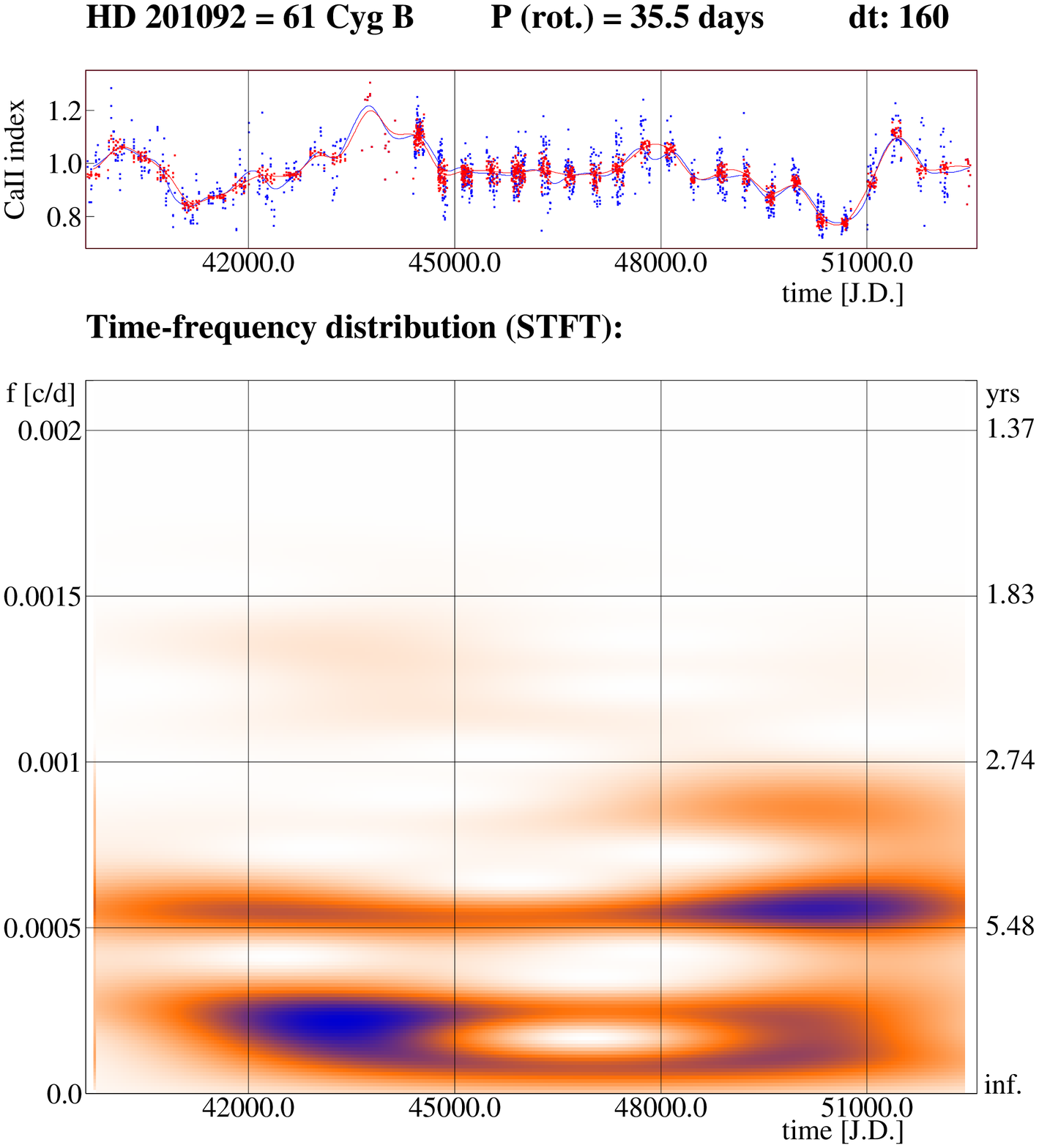}\hspace*{3mm}
\includegraphics[width=5.5cm]{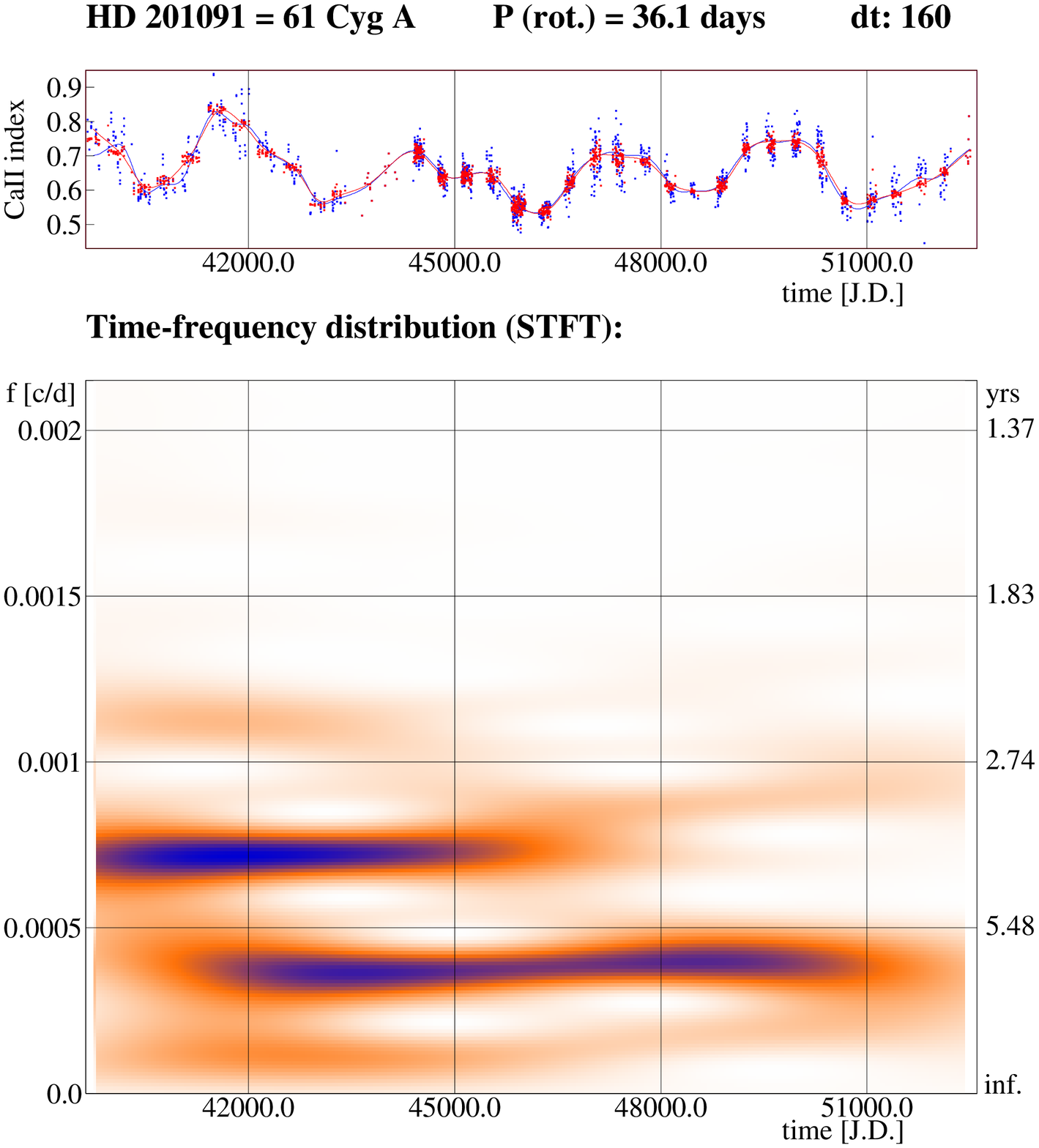}\hspace*{3mm}
\includegraphics[width=5.5cm]{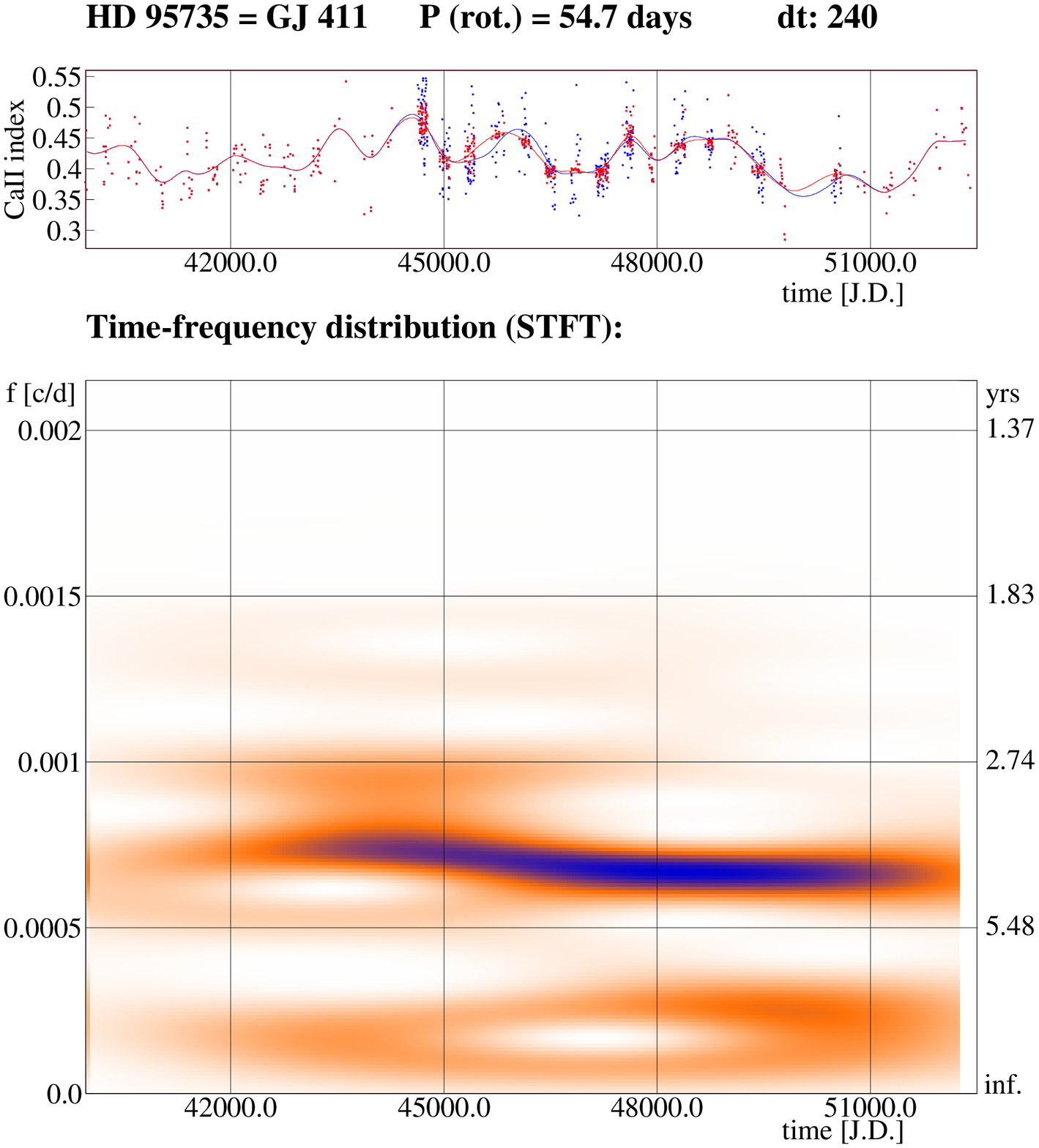}
\caption{The STFT of the Ca II sample, panels and colour codes are the same as in 
Fig.~\ref{F1}. Top: HD~131156A, HD~131156B, HD~100180, bottom: HD~201092, HD~201091, HD~95735. Details are in the text.}
    \label{F4}
\end{figure*}

{\bf HD~131156A = $\xi$ Boo A.} Variability on two
time scales are present. The shorter cycle is $\sim$5.5-yr and exhibits
a high amplitude in the beginning of the record;
the cycle length decreases to $\sim$3.9-yr at the record's end. A longer-period
variability is also apparent, with a characteristic timescale of  
$\sim$11 yr in the middle of the record. Baliunas et al. (\cite{sallie3}) marked
the decadal variability of the record as "var," which means {\it "significant variability without 
pronounced periodicity, on timescales longer, than 1-yr but much shorter than 25-yr..."}, and this is in accordance with our results for the record that is ten years longer; also, Baliunas et al. (\cite{sallie2}) find a long cycle of 13.2 years.

{\bf HD~131156B = $\xi$ Boo B.} Only one long-term periodicity is found, and 
only in the beginning of the record, approximately $\sim$4.3-yr, although
a secular change is also evident. Baliunas et al. (\cite{sallie3}) remark as "long", 
i.e., signifcant variability is found, which is longer than 25 years.

{\bf HD~100180 = 88 Leo.} We derive a more or less constant cycle of $\sim$3.5-yr,
which persists nearly through the entire record, but whose
amplitude is strong in the first half of the record and decreasing thereafter. Another variable cycle of $\sim$13.7-yr appears in the beginning of the record;
the period decreases to $\sim$8.6-yr by the end of the record, in agreement with the earlier results of Ol\'ah et al. (\cite{cycles3}). Our results in the beginning of the dataset are similar 
to those found by Baliunas et al. (\cite{sallie3}), who found two cycles,
$\sim$3.56 and $\sim$12.9-yr in the record shorter by ten years. Analyzing the 
full record, Frick et al. (\cite{frick}) found periodicities of $\sim$3.7 and $\sim$10.1-yr. The short cycle is consistent with our result, the $\sim$10.1-yr value may
be an averaged mean value of the of the long variable cycle we detect.

{\bf HD~201092 = 61 Cyg B.} The record for this star also exhibits two activity cycles: one is 
$\sim$4.7-yr and persists thoughout the record; the other has a timescale 
of $\sim$10-13 years. Both periodicities have variable amplitudes. 
The long cycle was estimated in previous studies as 11.4, 11.1 and 11.7-yr by Baliunas et al. 
(\cite{sallie3}), Frick et al. (\cite{frick}) and Baliunas et al. (\cite{sallie2}), respectively.

{\bf HD~201091 = 61 Cyg A.} The high amplitude cycle seen
in the record of this star has a mean length of 
6.7-yr, which slowly changes between 6.2 and 7.2-yr. A shorter, significant 
cycle is found in the first half of the record with a characteristic timescale 
of $\sim$3.6-yr. Probably because of the changing and double nature of the main 
cycles, previous determinations are slighly different but still consistent 
with our result: Baliunas et al. (\cite{sallie3}) derived 7.3-yr, Frick et 
al. (\cite{frick}) found 7.12-yr and finally Baliunas et al. (\cite{sallie2}) 
estimated 5.43-yr. However, the most interesting comparison to our 
results for HD~201091 is that of Frick et al. (\cite{frick1}, their Figure~5, bottom 
left panel, and Figure~6, lower panel). They derived a changing cycle between 
$\sim$6.7 and 7.9-yr, while the record at that time was about ten years shorter 
than now. Comparing Figure~5 of Frick et al. (\cite{frick1}) to Fig.~\ref{F4}, 
(bottom, middle), it is seen that the results are similar, though, 
because of our higher frequency resolution we could resolve two cycles in the 
beginning of the record.

{\bf HD~95735 = GJ~411.} The stronger cycle of this star is 
$\sim$3.9-yr, which is slightly shorter (3.4-yr) in the beginning of the 
record. A longer, 11-yr cycle is also present with a smaller amplitude. 
Baliunas et al. (\cite{sallie3}) classify the record as "var" (see the explanation 
at HD~131156A). The long cycle was estimated as $\sim$11.3-yr by Baliunas et al. 
(\cite{sallie2}). The rotational period found by Baliunas et al. 
(\cite{sallie2}) is 28.5-d and by us 54.7-d; the shorter 
result possibly arises from the non-sinusoidal nature of the rotational modulation 
caused by two distinct activity centres at spatially separated stellar longitudes.

%%%%%%%%%%%%%%%%%%%%%%%%%%%%%%%%%%%%
%%%%% D I S C U S S I O N %%%%%%%%%%
%%%%%%%%%%%%%%%%%%%%%%%%%%%%%%%%%%%%

\section{Discussion}\label{discussion}

\subsection{Cycle lengths}
%---------------------------------------  Table 6
\begin{table}[tbh]
  \caption{Derived activity cycles. }\label{T6}
  \begin{center}
  \leavevmode
  \footnotesize
  \begin{tabular}{lrl}
  \hline
  \noalign{\smallskip}
  star & \multicolumn{1}{c}{Rotation period$^a$}   &  \multicolumn{1}{c}{Cycle lengths} \\
       & \multicolumn{1}{c}{(days)}          &  \multicolumn{1}{c}{(years)} \\
  \noalign{\smallskip}
  \hline
  \noalign{\smallskip}
AB Dor    &  0.515 & 3.3, L \\
LQ Hya    &  1.601 & 2.5, 3.6, 7.0$\rightarrow$12.4 \\
V833 Tau  &  1.788 & 2.2, 5.2, 27-30, L \\
V410 Tau  &  1.872 & 4.1$\leftarrow$5.2, 6.5$\rightarrow$6.8 \\
EI Eri    &  1.947 & 2.9$\leftrightarrow$3.1, 4.1$\leftrightarrow$4.9, 14.0 \\
FK Com    &  2.400 & 4.5$\leftrightarrow$6.1, L \\
V711 Tau  &  2.838 & 3.3, 5.4, 8.8$\leftarrow$17.9 \\
UZ Lib    &  4.768 & 3.1$\leftarrow$4.3, L \\
UX Ari    &  6.437 & 3.6$\leftarrow$4.6, L \\
HU Vir    & 10.388 & 5.7, L \\
IL Hya    & 12.905 & 2.7$\rightarrow$3.0, 4.4, L \\
XX Tri    & 23.969 & 3.8, L \\
HK Lac    & 24.428 & 5.4$\rightarrow$5.9, 10.0$\rightarrow$13.3, L \\
IM Peg    & 24.649 & 9.0, L \\
\hline
HD 131156A &  6.25   & 3.9$\leftarrow$5.5, 11.0 \\	
HD 131156B & 11.05   & 4.3, L \\
HD 100180  & 14.32   & 3.5, 8.6$\leftarrow$13.7 \\
HD 201092  & 35.53   & 4.7, 11.5 \\
HD 201091  & 36.06   & 3.6, 6.2$\leftrightarrow$7.2 \\
HD 95735   & 54.7    & 3.4$\leftrightarrow$3.9, 11.0 \\
  \noalign{\smallskip}
  \hline
  \end{tabular}
  \end{center}
The cycle lengths, given in years, are listed in order of increasing value. For changing cycle periods, arrows show the direction of the change between the extreme values.`L' denotes multi-decadal variations.\\
$^a$ in the case of a synchronized binary, the orbital period is given\\
\end{table}
%---------------------------------------  Table 7
\onltab{7}{
\begin{table}[tbh]
  \caption{Derived activity cycles. }\label{T7}
  \begin{center}
  \leavevmode
  \footnotesize
  \begin{tabular}{lrl}
  \hline
  \noalign{\smallskip}
  star & \multicolumn{1}{c}{Rotational period$^a$}   &  \multicolumn{1}{c}{Cycle lengths} \\
       & \multicolumn{1}{c}{(days)}          &    \multicolumn{1}{c}{(days)} \\
  \noalign{\smallskip}
  \hline
  \noalign{\smallskip}
AB Dor    &  0.515 & 1200, L \\
LQ Hya    &  1.601 & 909, 1316, 2571$\rightarrow$4545 \\
V833 Tau  &  1.788 & 794, 1898, 10000, L \\
V410 Tau  &  1.872 & 1493$\leftarrow$1887, 2381$\rightarrow$2500 \\
EI Eri    &  1.947 & 1066$\leftrightarrow$1136, 1506$\leftrightarrow$1786, 5100 
\\
FK Com    &  2.400 & 1639$\leftrightarrow$2222, L \\
V711 Tau  &  2.838 & 1200, 1957, 3195$\leftarrow$6536 \\
UZ Lib    &  4.768 & 1130$\leftarrow$1560, L \\
UX Ari    &  6.437 & 1299$\leftarrow$1667, L\\
HU Vir    & 10.388 & 2100, L \\
IL Hya    & 12.905 & 993$\rightarrow$1111, 1618, L\\
XX Tri    & 23.969 & 1400, L \\
HK Lac    & 24.428 & 1957$\rightarrow$2150, 3650$\rightarrow$4854, L \\
IM Peg    & 24.649 & 3298, L \\
\hline
HD 131156A &  6.25   & 1429$\leftarrow$2000, 4000 \\	
HD 131156B & 11.1   & 1563, L \\
HD 100180  & 14.3   & 1285, 3125$\leftarrow$5000 \\
HD 201092  & 35.5   & 1724, 4200 \\
HD 201091  & 36.1   & 1320, 2262$\leftrightarrow$2632 \\
HD 95735   & 54.7    & 1250$\leftrightarrow$1429, 4000 \\
  \noalign{\smallskip}
  \hline
  \end{tabular}
  \end{center}
The cycle 
lengths, given in days, are listed in order of increasing value. For changing cycle periods, arrows show the 
direction of the change between the extreme values.`L' denotes multi-decadal variations.\\
$^a$ in the case of a synchronized binary, the orbital period is given\\
\end{table}
}

The results of our analysis are summarised in Table~\ref{T6}, where the cycle 
lengths (yr, and in the electronic Table~\ref{T7}, in d) are listed in order of increasing value. 
In cases of changing cycle periods, arrows show the 
direction of the change between the extreme values (increasing, decreasing, or 
both). An `L' denotes a multi-decadal variation, which may occur in addition to the detected 
cycle(s); however, a timescale cannot be given because of the insufficient length of the record to resolve the longer period.

Most of the stars show multiple cycles. Those stars for which we list only one 
cycle usually show variability on longer timescales. The lengths of the 
records are too short to verify additional, longer cycles for AB~Dor, HU~Vir, 
XX~Tri, IM~Peg and HD~131156B (see Table~\ref{T6}).

The cycle lengths are not strictly constant or randomly variable, but show {\it 
systematic} changes during the period of observations. We see two kinds of 
systematically changing cycles. The first is characterized by a cycle length that oscillates 
between a lower and an upper value, and the second by a cycle length exhibiting a 
secular variation, i.e., an increase or a decrease. However, these 
continuously increasing or decreasing cycle lengths may reverse over a longer observational period than currently available.

In Paper I we studied the multi-decadal variability of the solar Schwabe and 
Gleissberg cycles during the last 250 years from Sunspot Number records. In Figure 2 (bottom panel) of Paper I
we used the same method (STFT) as 
for the stars in the present paper. The results look similar -- i.e., 
for the Sun, one cycle (Schwabe) varies between limits, while the longer one (Gleissberg) 
continually increases. However, the results for a longer time displayed on Figure 8 in Paper I indicated that the 
Gleissberg cycle of the Sun also varies between limits on a scale of $\sim$50-200-yr, 
but to confirm the variability of the Gleissberg cycle required a longer, 500-yr record. By analogy from the analysis of the longer
solar record, the presence of a long-term trend may suggest an increasing or decreasing multidecadal cycle that is presently unresolved in the stellar records of short duration. 

Another feature of interest seen in our analysis of cycle lengths in stellar
records is that of the dominant (highest amplitude) cycle rapidly
switching to another one. In EI~Eri in the beginning of the dataset 
the long, $\sim$14-yr cycle is dominant, and the short cycle of about 4.5-yr is only weakly present. Around JD 2447000 the amplitude of this short cycle increases considerably, and that high-amplitude cycle persists for more than 12-yr ($\sim$3 short cycles). Later the longer cycle again dominates. (The beginning is uncertain because of sparse data, but the end of
the record is well sampled.) A similar pattern is seen in LQ~Hya.

Interesting feature is found in V410~Tau, EI~Eri and UX~Ari: two cycles in each of them
which are not harmonics of each other, are changing parallel. This finding points 
towards the common physical origin of these cycles, i.e., that both cycles are 
produced by the same dynamo.

\subsection{Relations}

\begin{figure}[tbh]
\begin{center}
\includegraphics[width=9cm]{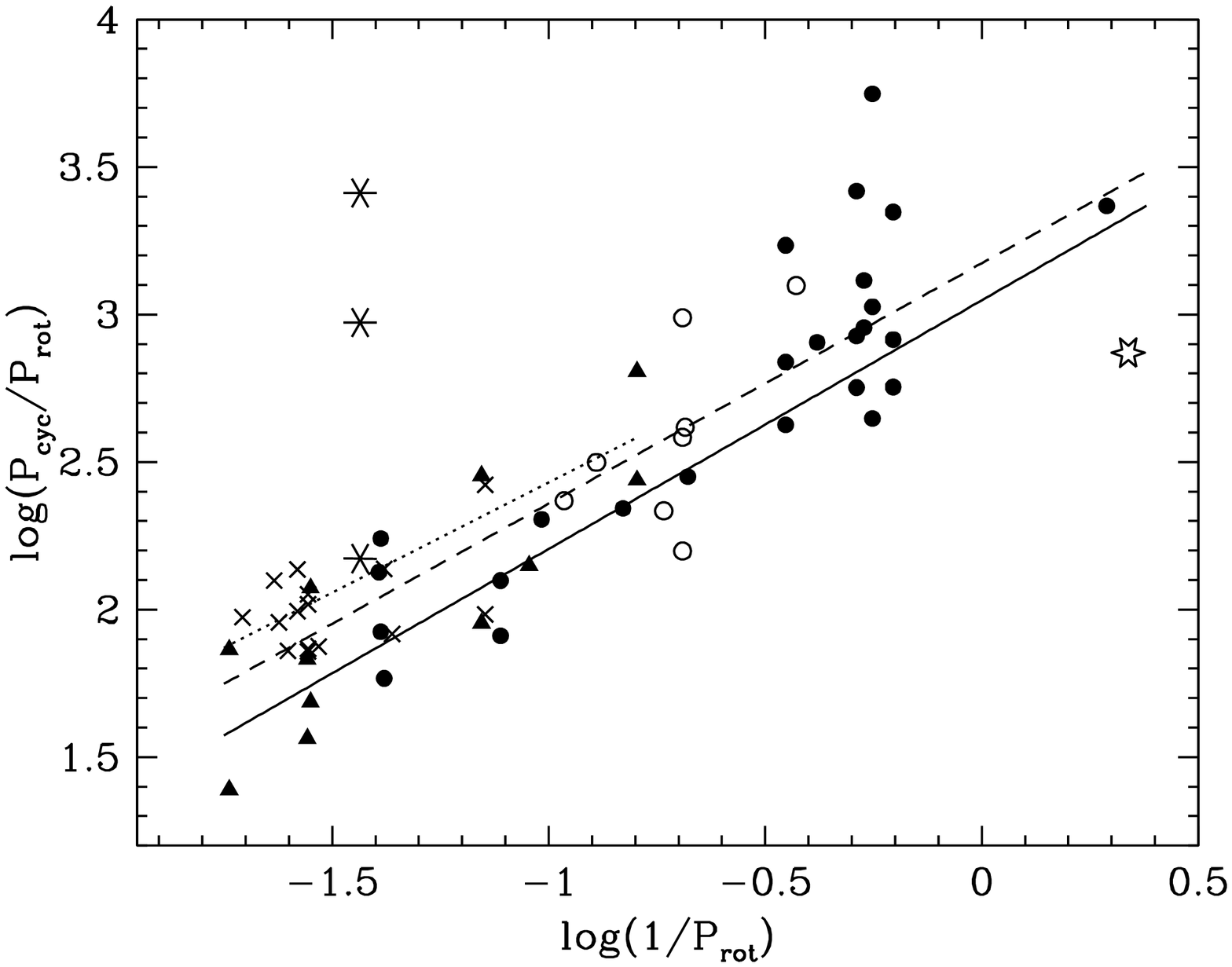}
\caption{The quantity $log(P_{cyc}/P_{rot})$ vs.$log(1/P_{rot})$, as in Baliunas 
et al. (\cite{sallie4}). The plot includes the results of the present 
investigation, and also results from the literature: Messina \& Guinan 
(\cite{messina}, circles), and Frick et al. (\cite{frick}, crosses). The results 
from the current photometric investigation are marked by filled dots, and filled 
triangles denote the results from the Ca II records (HD~95735 is excluded from the fits 
and the mean of its changing cycles are plotted). The different cycles of the 
Sun are represented by stars. A hexagon in the far right shows the position of 
the fast rotator EY~Dra (excluded from the fits). The dotted line shows the original 
relation from Baliunas et al. (\cite{sallie4}), dashed line is the fit to all 
results, including those from the literature, and the solid line is the relation 
for the shortest cycle lengths determined by us.}
    \label{F5}
\end{center}
\end{figure}

\begin{figure}[tbh]
\begin{center}
\includegraphics[width=9cm]{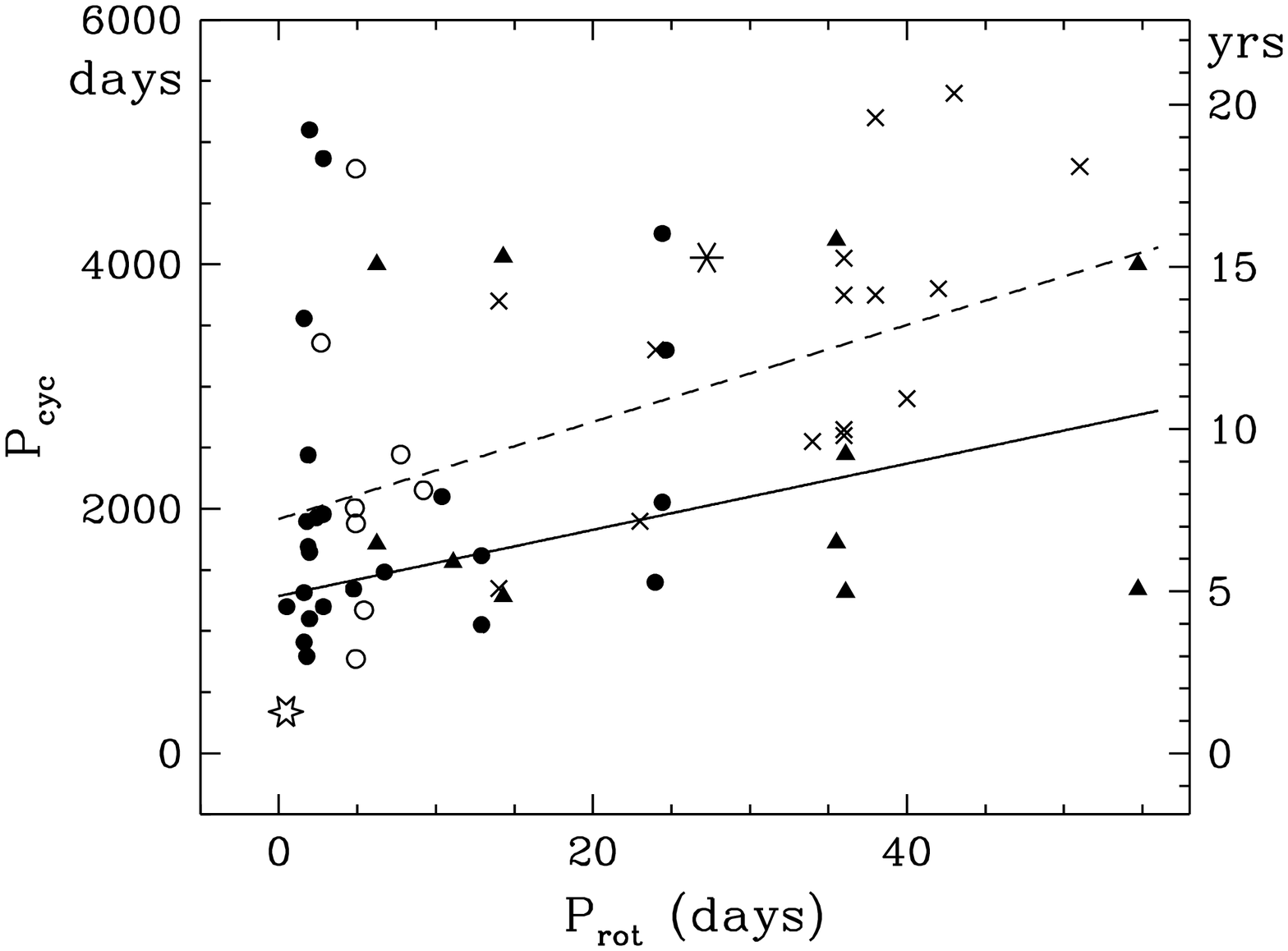}
\caption{The rotational period-cycle period diagram, symbols are as in 
Fig.~\ref{F5}. The dashed line is the fit to all results including the ones from the 
literature, and solid line is the relation for the shortest cycle lengths 
determined by us.}
    \label{F6}
\end{center}
\end{figure}

The cycle lengths of two stars were excluded from the further analysis. The 
first is the shortest period, ultrafast rotating ($P_{rot}=0.459$ d), M1-2 
dwarf star EY~Dra, which might show a cycle of around 1 year from observations 
made every two months with gaps of only 4 months (see Vida 
\cite{krisztian}), but that cycle is uncertain. The other is the  star with the longest rotation period in our sample, HD~95735 ($P_{rot}=54.7$ d), a dwarf of spectral type M2, which is
rotating $\sim$120 times slower than EY~Dra, but has the same spectral 
type. Those two stars are the only M dwarfs which have reported activity cycles, and
have the shortest (EY~Dra) and longest (HD~95735) rotation periods, which
are plotted in Fig.~\ref{F5} and Fig.~\ref{F6}, but are excluded from the fits. Their internal structure may differ significantly from the other stars in the sample.

Graphical representation of the results are given in Fig.~\ref{F5} on a 
$log-log$ scale of $P_{cyc}/P_{rot}$  vs. $1/P_{rot}$, as in the 
paper of Baliunas et al. (\cite{sallie4}, Figure~1). The cycle period is expected 
to scale $\sim D^{\iota}$ where ${\iota}$ is the slope of the relation and $D$ 
is the dynamo number. The slope for the results in Baliunas et al. 
(\cite{sallie4}) was 0.74 (dotted line in Fig.~\ref{F5}), and we get $0.81\pm0.05$ 
for all the results from the present paper and from Messina \& Guinan 
(\cite{messina}), and Frick et al. (\cite{frick}). The fit to the shortest cycles 
determined by us gives a slope of $0.84\pm0.06$.

The $log-log$ representation of the cycle period normalised by the rotation period 
gives a direct relation to the dynamo number. In comparison, 
the direct representation of the cycle vs. rotational periods in Fig.~\ref{F6} has 
information about the tightness of the relation that is in Fig.~\ref{F5}
suppressed a little by the normalization and the scale. All symbols of Fig.~\ref{F6} are the same as in Fig.~\ref{F5}. The Gleissberg timescale of the Sun is not plotted, because even 
its shortest value (50-yr, i.e., about 18000-d) would compress the figure 
along the y-axis. The slope of the relation for all the results of the present 
paper together with the values from the literature is $39.7\pm10.6$, whereas for 
the newly determined shortest cycles is $27.1\pm12.1$, i.e., the relation is 
significant for all the data considered together, and marginal when only the shortest cycles are considered. 

However, an important warning should be recalled from Paper I: cycle periods of 
1-2-yr recovered from records with approximately yearly gaps are highly uncertain if not insignificant. Hence, the lower end of the diagrams in  Fig.~\ref{F5} and 
Fig.~\ref{F6}, i.e., at the shorter rotational periods, the shortest cycles 
simply cannot be recovered with certainty, thereby distorting the fitted relations.

Finally, we note that the stars studied in this paper and those from the 
literature that are plotted together in Fig.~\ref{F5} and Fig.~\ref{F6} 
represent a wide range of objects in the context of spectral types, characteristics of the 
observational data (photometry, spectroscopy, lengths of the 
records), and that the sample consists of both single stars and stars in binary 
systems. Therefore, while we find a general trend between the rotational 
periods and cycle timescales, a much tighter correlation cannot be expected.

%%%%%%%%%%%%%%%%%%%%%%%%%%%%%%%%%%%%
%%%%% S U M M A R Y %%%%%%%%%%%%%%%%
%%%%%%%%%%%%%%%%%%%%%%%%%%%%%%%%%%%%

\section{Summary}

We analysed continuous photometric and spectroscopic records
on multi-decadal periods of 20 active, cool stars 
using Short-Term Fourier Transform and studied the patterns of their activity 
cycles. We found that

\begin{itemize}
\item 15 stars show definite, multiple cycles, the others probably do, 
\item most of the cycle lengths are found to change systematically,
\item in three cases two cycles are changing parallel,
\item and the relation between the rotational periods and cycle lengths is 
confirmed
\end{itemize}

for the inhomogeneous set of active stars consisting of dwarfs and giants, 
single and binary systems.

Moss et al. (\cite{moss_et_al}) depict a complicated picture of symmetry 
properties derived from models of stellar magnetic fields, while expecting that
properties of observed stellar cycles may offer progress on the problem. The present paper gives twenty different examples (21 together with the Sun from Paper I) of multiple cycles 
that can arise from {\it "dynamos operating in significantly supercritical 
regimes"}. The observed modulations in amplitude and cycle lengths may originate 
from cyclic quadrupolar components superimposed on a dipole cycle, as 
suggested by Moss et al. (\cite{moss_et_al}). The {\it predictability} of the 
cycle patterns at the moment is very limited: we are able to describe from 
observations with mathematical tools what happened in the past, but for the future only trends and tendencies can be suggested. Further detailed theory is necessary to model the complicated cycle structures we observe in stars and in the Sun. Thorough modeling the observed cycle 
patterns may be for the present the only way to predict with success the future patterns of the Sun and the stars.

% =================================== a c k n o w l e d g e m e n t s

\begin{acknowledgements}
K.O. acknowledges support  from the Hungarian Research Grants OTKA T-048961 and 
T-068626. S.B. acknowledges support from
JPL 12700064, Smithsonian Institution Restricted
Endowment funds and NASA NNX07AI356. KGS appreciates the continuous support of the Vienna-AIP 
APTs in southern Arizona through the State of Brandenburg MWFK.
\end{acknowledgements}

% =================================== R e f s

% -----------------------------------------------------------------------
%
%   A P P E N D I X  -  online material
%
% -----------------------------------------------------------------------

\Online

\end{document}